\documentclass[aps,prc,superscriptaddress,twocolumn]{revtex4-1}

\usepackage[dvipdfmx]{graphicx}
\usepackage{dcolumn}% Align table columns on decimal point
\usepackage{bm}% bold math
\usepackage{rotating}
\usepackage{amssymb}
\usepackage{color}
\begin{document}

\title{In-gas-cell laser spectroscopy for magnetic dipole moment of $^{199}$Pt toward $N=$ 126}

\author{Y. Hirayama}
\email{yoshikazu.hirayama@kek.jp}
\affiliation{Wako Nuclear Science Center (WNSC), Institute of Particle and Nuclear Studies (IPNS), High Energy Accelerator Research Organization (KEK), Wako, Saitama 351-0198, Japan}
\author{M. Mukai}
\affiliation{University of Tsukuba, Tsukuba, Ibaraki 305-0006, Japan}
\affiliation{Wako Nuclear Science Center (WNSC), Institute of Particle and Nuclear Studies (IPNS), High Energy Accelerator Research Organization (KEK), Wako, Saitama 351-0198, Japan}
\affiliation{Nishina Center for Accelerator-Based Science, RIKEN, Wako, Saitama 351-0198, Japan}
\author{Y.X. Watanabe}
\affiliation{Wako Nuclear Science Center (WNSC), Institute of Particle and Nuclear Studies (IPNS), High Energy Accelerator Research Organization (KEK), Wako, Saitama 351-0198, Japan}
\author{S.C. Jeong}
\affiliation{Rare Isotope Science Project, Institute for Basic Science (IBS), Daejeon 305-811, Republic of Korea}
\affiliation{Wako Nuclear Science Center (WNSC), Institute of Particle and Nuclear Studies (IPNS), High Energy Accelerator Research Organization (KEK), Wako, Saitama 351-0198, Japan}
\author{H.S. Jung}
\affiliation{Wako Nuclear Science Center (WNSC), Institute of Particle and Nuclear Studies (IPNS), High Energy Accelerator Research Organization (KEK), Wako, Saitama 351-0198, Japan}
\author{Y. Kakiguchi}
\affiliation{Wako Nuclear Science Center (WNSC), Institute of Particle
  and Nuclear Studies (IPNS), High Energy Accelerator Research
  Organization (KEK), Wako, Saitama 351-0198, Japan}
\author{\\S. Kimura}
\affiliation{University of Tsukuba, Tsukuba, Ibaraki 305-0006, Japan}
\affiliation{Wako Nuclear Science Center (WNSC), Institute of Particle and Nuclear Studies (IPNS), High Energy Accelerator Research Organization (KEK), Wako, Saitama 351-0198, Japan}
\affiliation{Nishina Center for Accelerator-Based Science, RIKEN, Wako, Saitama 351-0198, Japan}
\author{J.Y. Moon}
\affiliation{Rare Isotope Science Project, Institute for Basic Science (IBS), Daejeon 305-811, Republic of Korea}
\author{M. Oyaizu}
\affiliation{Wako Nuclear Science Center (WNSC), Institute of Particle and Nuclear Studies (IPNS), High Energy Accelerator Research Organization (KEK), Wako, Saitama 351-0198, Japan}
\author{J.H. Park}
\affiliation{Rare Isotope Science Project, Institute for Basic Science (IBS), Daejeon 305-811, Republic of Korea}
\author{P. Schury}
\affiliation{Wako Nuclear Science Center (WNSC), Institute of Particle and Nuclear Studies (IPNS), High Energy Accelerator Research Organization (KEK), Wako, Saitama 351-0198, Japan}
\author{M. Wada}
\affiliation{Wako Nuclear Science Center (WNSC), Institute of Particle and Nuclear Studies (IPNS), High Energy Accelerator Research Organization (KEK), Wako, Saitama 351-0198, Japan}
\affiliation{Nishina Center for Accelerator-Based Science, RIKEN, Wako, Saitama 351-0198, Japan}
\author{H. Miyatake}
\affiliation{Wako Nuclear Science Center (WNSC), Institute of Particle and Nuclear Studies (IPNS), High Energy Accelerator Research Organization (KEK), Wako, Saitama 351-0198, Japan}

\date{\today}

\begin{abstract}
Magnetic dipole moment and mean-square charge radius of $^{199}$Pt
($I^{\pi}=$ 5/2$^-$) have been evaluated for the first time from the
investigation of the hyperfine splitting of the $\lambda_1=$ 248.792
nm transition by in-gas-cell laser ionization spectroscopy. 
Neutron-rich nucleus $^{199}$Pt was produced by
multi-nucleon transfer reaction at the KISS where the nuclear
spectroscopy in the vicinity of $N=$ 126 is planed from the aspect of
an astrophysical interest as well as the nuclear structure.
Measured magnetic dipole moment $+$0.63(13)$\mu_{\rm N}$ is consistent
with the systematics of those of nuclei with $I^{\pi}=$ 5/2$^-$.
The deformation parameter $|<\beta_2^2>^{1/2}|$ evaluated from
the isotope shift indicates the gradual shape change to spherical shape of
platinum isotopes with increasing neutron number toward $N=$ 126.
\end{abstract}
\pacs{21.10.Ky \sep 21.10.Ft \sep 27.80.+w \sep 42.62.Fi}

\maketitle

\section{INTRODUCTION}\label{sec-intro}
Allowed Gamow-Teller (GT) and first-forbidden (FF) beta-decay
transitions compete in the nuclei around $N$ = 126, depending on the
shell evolution at increasing the number of neutrons.
The proton orbit of the neutron-rich nuclei around $N=$ 126 would be
$\pi(0h_{11/2})$,  therefore, the GT transition from $\nu(0h_{9/2})$
to $\pi(0h_{11/2})$, and the FF transition from $\nu(0i_{13/2})$ to
$\pi(0h_{11/2})$ would be competitive. 
This makes it difficult to predict nuclear properties such as the
half-life and nuclear mass in the neutron-rich nuclei around $N=$ 126
by theoretical nuclear models \cite{BOR03,MOL03,KOU05}.
Therefore, the half-lives are largely deviated by one order of
magnitude \cite{MOL97,ENG99,PIN01,LAN03,KUTY}.
The half-lives play an important role to investigate an
astrophysical environment for the formation of the third peak in an
observed solar r-process abundance pattern \cite{BUR57,MUM16}.
Because the abundance peak height around $A=$ 195 is proportional to
the half-lives, and the peak position and width drastically change
according to the half-lives.
The experimental nuclear spectroscopy are desirable for eliminating
the uncertainty and improving the predictions by nuclear models taking
into account the FF transition.

The study of nuclear structure toward $N=$ 126 through electromagnetic
moments has been intensively performed by various techniques suitable
for the nuclear properties such as half-life, level energy and decay modes \cite{STO11}.
The nuclear wave-function and shape can be evaluated from the measured magnetic
dipole moment ($\mu$) and quadrupole moment ($Q$), respectively.
Especially, the understanding of the wave function on the ground state in
this region is essential to provide the accurate prediction of GT and FF transition
strength, namely, $\beta$-decaying half-lives.
However, the difficulty of the production of the nuclei in the
vicinity of $N=$ 126 causes the lack of the electromagnetic moments.
In order to break through the situation, we have started KISS project
\cite{JEO10}. 
In the project, the nuclei around $N=$ 126 are produced by
multi-nucleon transfer (MNT) reaction \cite{DAS94} of $^{136}$Xe beam
and $^{198}$Pt target system \cite{WAT15}, and are selected by an
argon gas-cell based laser ion source combined with on-line isotope
separator (KISS) \cite{HIR15,HIR16}.
The electromagnetic moments can be measured by in-gas-cell and -gas-jet
laser ionization spectroscopy (IGLIS) \cite{COC10,FER14,KUD13}.

By taking the advantage of the MNT reaction combined with the KISS,
we have started to produce the neutron-rich isotopes of the refractory elements
such as Pt, Ir, Os, Re, W and Ta, and perform the nuclear spectroscopy systematically.
The search for the laser resonance ionization scheme of the refractory
elements has been in progress \cite{HIR14,MUK15,MUK16}.
As a first step,  the measurement of the magnetic dipole moment of
$^{199}$Pt ($Z=$ 78, $N=$ 121, $I^{\pi}=$ 5/2$^-$, and $t_{1/2}=$
30.8(2) min.) \cite{TOI} was performed by in-gas-cell laser ionization
spectroscopy.
Nuclear structure of platinum isotopes with 178 $\leq A \leq$ 198 have
been well studied through the measured $\mu$ and $Q$, isotope shift
(IS) and charge radius by the laser spectroscopy \cite{LEE88,DUO89,HIL92,KIL92,BLA99}.
In the lighter, neutron-deficient, region of 178$\leq A \leq$183,
shape coexistence was found and discussed from the IS \cite{BLA99}.
On the other hand, in the heavier region of 183$\leq A \leq$198,
the prolate (183$\leq A \leq$188), triaxial shape (190$\leq A
\leq$194) and oblate (196$\leq A \leq$ 198) shapes were discussed and compared
with theoretical models \cite{LEE88,DUO89,HIL92,KIL92}.
In the more neutron-rich region, spherical shape would become
dominant toward the shell-closure of $N=$ 126 in the same way as lead
($Z=$ 82) and mercury ($Z=$ 80) isotopes \cite{WIT07} predicted by the
droplet model \cite{WIL83} with $\beta_{\rm 2}=$ 0.
In this paper, we report the experimental details and the results for the
in-gas-cell laser ionization spectroscopy to determine the magnetic
dipole moment of neutron-rich $^{199}$Pt.

\section{EXPERIMENT}
\subsection{Principle of laser spectroscopy for deducing nuclear electromagnetic moments}
Laser spectroscopy is powerful tool to determine nuclear
electromagnetic moments through the investigation of hyperfine levels
governed by quantum number $F$.
The possible $F$ values are in the range of $|I - J| \leq F \leq I +
J$ where $I$ and $J$ are nuclear and atomic spins, respectively.
According to $F$ values, if $I >$ 1/2, the degenerated atomic energy
levels are resolved and changed by $\Delta E$ which is denoted as following;
\begin{eqnarray}
\Delta E &=& \frac{A}{2} \times K \nonumber \\[1mm]
         & & + \frac{B}{2} \times
\frac{3K(K+1)-2I(I+1)2J(J+1)}{2I(2I-1)2J(2J-1)}, \label{eq:fitting1} \\[1mm]
      K&=&F(F+1)-I(I+1)-J(J+1). \nonumber
\end{eqnarray}
Here, $A$ and $B$ are the magnetic-dipole and
electric-quadrupole hyperfine coupling constants, respectively. 
These factors are represented to be
\begin{equation}
A = \frac{\mu H_{JJ}({\rm 0}) }{IJ} \label{eq:a-factor} 
\end{equation}
and\\
\begin{equation}
B = eQ \phi_{JJ}({\rm 0})\hspace{1em} (I, J > 1/2), \label{eq:b-factor}
\end{equation}
respectively.
The $A$ and $B$ factors are related to the nuclear magnetic dipole
moment $\mu$ and nuclear electric quadrupole moment $Q$, respectively.
$H_{JJ}({\rm 0})$ and $\phi_{JJ}({\rm 0})$ are the magnetic field and
the electric field gradient, respectively, induced by the atomic electrons at
the position of the nucleus.
$H_{JJ}({\rm 0})$ and $\phi_{JJ}({\rm 0})$ are specific to each atomic
state, and are common for isotopes.
Therefore, nuclear electromagnetic moments can be evaluated from the
measured $A$ and $B$ factors by using the known $A'$ and $B'$ factors,
$I'$, $\mu'$, and $Q'$ of specific isotope (generally stable isotope) as follows;
\begin{equation}
\mu = \frac{I}{I'} \frac{A}{A'} \mu', \label{eq:gmom} 
\end{equation}
and\\
\begin{equation}
Q = \frac{B}{B'} Q' \hspace{1em} (I, J > 1/2). \label{eq:qmom}
\end{equation}

The atomic energy changes of $\Delta E_{\rm gs}$ and $\Delta E_{\rm
  ex}$ at the ground and excited states occur, respectively, as a
result of hyperfine interaction between the nucleus and atomic electrons.
The transition frequency between the hyperfine levels of the ground
and excited states shifts to be 
\begin{equation}
\Delta \nu_{\rm i} = \Delta E_{\rm  i:ex} - \Delta E_{\rm i:gs} \label{peak-pos}
\end{equation}
relative to the center of gravity of the fine-structure transition-frequency $\nu_0$.
From the laser spectroscopy for the measurement of the hyperfine
splitting expressed by the $\Delta \nu_{\rm i}$ values, the hyperfine
coupling constants $A$ and $B$, namely, electromagnetic moments $\mu$
and $Q$, can be evaluated.

\subsection{Experimental details}
\subsubsection{KISS}
The experiment was performed using KEK isotope separation system
(KISS) \cite{HIR15,HIR16}, argon gas-cell based laser ion source combined
with on-line isotope separator, installed in RIKEN Nishina center.
Primary beam $^{136}$Xe (10.75 MeV/$A$, 20 particle-nA), accelerated by
RIKEN Ring Cyclotron, impinged on an enriched $^{198}$Pt (purity
91.63\%) target with a thickness of 11 mg/cm$^2$.
Unstable nucleus $^{199}$Pt was produced by multi-nucleon transfer
reaction \cite{WAT15}, and stable nucleus $^{198}$Pt was also emitted
from the target by an elastic reaction with $^{136}$Xe.
At an off-line test to study the systematics of the laser ionization
for platinum isotopes, the stable platinum isotopes of
$^{192, 194,195,196, 198}$Pt were produced from the resistive heating of a
natural platinum filament placed in the gas cell.

The platinum isotopes were accumulated, thermalized and neutralized in
the argon gas cell with a pressure of 88 kPa, whose design was
optimized for the high efficient gas-flow transport \cite{HIR15}.
The isotopes were re-ionized in the gas cell by two-step resonant
laser ionization technique, and the atomic number $Z$ can be selected.
The laser-produced singly-charged ($q=$ 1) platinum ions were
extracted with an energy of 20 kV, and their mass-to-charge ratio
($A/q$) was selected by a dipole magnet.
Finally, one kind of isotope was transported to the detector station
placed at the neighboring experimental hall to the gas cell system.

The detector station includes a tape transport device for avoiding the
radioactivities in the decay chain of separated nuclides under pulsed
beam operation of the KISS.
The radioactive isotope was implanted on an aluminized Mylar tape
surrounded by three sets of two-layered plastic-scintillator telescopes
with 50\% detection efficiency \cite{KIM16}. 
The tape was moved to remove unwanted $\beta$-rays at the end of each
measurement cycle.
The ions of the stable isotopes were counted by using a Channeltron detector.

\subsubsection{Laser ionization}
\begin{figure}
    \begin{center}
      \resizebox{0.45\textwidth}{!}{
        \includegraphics{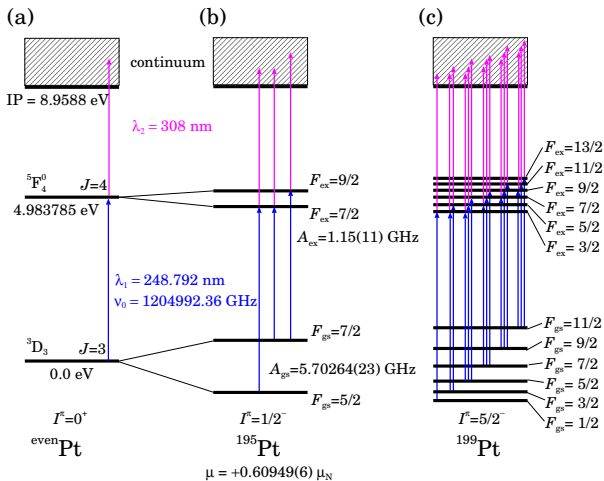}
      }
     \caption{Ionization scheme and hyperfine structure of platinum
       isotopes. (a) in the case of $^{\rm even}$Pt with $I^{\pi}=$ 0$^+$, there is no
       hyperfine splitting. (b) in the case of $^{195}$Pt with
       $I^{\pi}=$ 1/2$^-$ and $\mu=$ $+$0.60949(6) $\mu_{\rm N}$,
       the hyperfine splitting is governed by only $A_{\rm gs}$ and
       $A_{\rm ex}$ factors, and there are three transitions.
       (c) in the case of $^{199}$Pt with $I^{\pi}=$ 5/2$^-$,
       the hyperfine splitting of $F_{\rm gs}=$ 1/2--11/2 and $F_{\rm
         ex}=$ 3/2-13/2 is governed by $A$ and $B$ factors of the
       ground and excited states, and there are 15 transitions.
     }\label{HFS}
    \end{center}
\end{figure}

Figure \ref{HFS} shows the ionization scheme and hyperfine structure
of platinum isotopes.
Tunable dye laser (Radiant Dye Laser; NarrowScan) pumped by
XeCl:excimer laser ($\lambda=$ 308 nm, Lambda Physik LPX240i)
was utilized for the first step excitation ($\lambda_{\rm 1} =$ 248.792
nm) with laser power 70 $\mu$J/pulse.
Another XeCl excimer laser of $\lambda_{\rm 2} =$ 308 nm with laser
power 9 mJ/pulse was used as the ionization transition to continuum
above the ionization potential (IP). 
In the present laser ionization spectroscopy, the transition between
the ground state 5d$^9$6s$^3$D$_3$ and the excited state
5d$^8$6s6p$^5$F$^0_4$ was studied by scanning the wavelength
$\lambda_{\rm 1}$.
The systematic study concerning the gas-pressure effects was
performed by using stable platinum isotopes with mass number $A=$ 192,
194, 195, 196, and 198.
Pressure broadening and shift in the range of 11--88 kPa were studied
to be 39(3) MHz/kPa and $-$22(2) MHz/kPa, respectively. 
Isotope shifts for the stable platinum isotopes were also studied. 

The excimer lasers were operated at a repetition rate of up to 200 Hz
and were synchronized by master trigger signals from a function generator.
To produce a UV laser beam as first step light, radiation
delivered by the NarrowScan was frequency doubled using a barium
borate (BBO) crystal placed in a second-harmonic generator.
The typical linewidth and pulse width of the dye lasers were 3.4 GHz
and 15 ns, and the wavelength was monitored by a wave-meter WS6 (HighFinesse).
The 10\%  of UV laser power through a laser beam splitter was
monitored during the experiment.
The distance between the laser system and the gas cell is about 15 m.
Both laser beams with the size of about 8--10 mm in diameter are
transported to the gas cell with a small angle, and are overlapped in
the gas cell for resonance ionization, spatially and temporally.
The laser spots through the gas cell was checked by using a camera in
order to keep the spatial overlap by adjusting the mirrors with actuators.
The timing signals from photodetectors were measured by an
oscilloscope in order to maintain temporal overlap of the two lasers
by adjusting the master trigger signal in 1--ns steps.
The spatial and temporal overlapping were always monitored and
adjusted if necessary.

\section{EXPERIMENTAL RESULTS AND DISCUSSION}
\subsection{Identification of $^{199}$Pt nucleus}\label{id}
\begin{figure}
    \begin{center}
      \resizebox{0.45\textwidth}{!}{
        \includegraphics{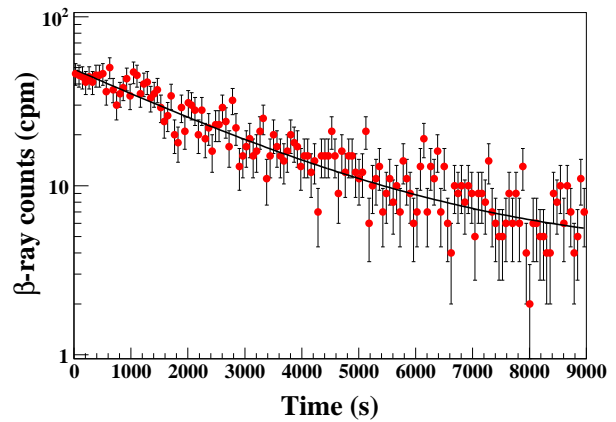}
      }
        \caption{Measured decay curve of $^{199}$Pt nucleus, which was
          used  for the identification of the $^{199}$Pt nucleus extracted from the
          KISS.
        }\label{199pt_life}
    \end{center}
\end{figure}

In the experiment, the laser-ionized $^{198}$Pt$^+$ was used to adjust
the beam line optics at first.
Then, the extraction of laser-ionized radioactive isotope $^{199}$Pt$^+$ was
performed only by changing the magnetic field of the dipole magnet.
Identification of $^{199}$Pt$^+$ nucleus ($t_{\rm 1/2}=$ 30.8 $\pm$
0.2 min. \cite{TOI}) was done by measuring the $\beta$-decay half-life
$t_{\rm 1/2}$ as shown in Fig. \ref{199pt_life}.
After 30 min. irradiation of $^{199}$Pt$^+$ beam on the tape, the
decay curve was measured during 2.5 hours ( $\approx$ 5 $\times t_{\rm 1/2}$).
Solid line in Fig. \ref{199pt_life} is the best fit result obtained by
using the fitting routine of MINUIT code \cite{MIN94}, and the fitting function consists of single
exponential function (free parameters of amplitude and $t_{\rm 1/2}$)
and constant background.
Measured half-life $t_{\rm 1/2}=$ 31.3 $\pm$ 1.5 min. was in good
agreement with the literature value $t_{\rm 1/2}=$ 30.8(2) min.
This result indicates that measured $\beta$-rays were emitted from
only $^{199}$Pt$^+$ extracted from the gas cell.

\subsection{Extraction yield of $^{199}$Pt nucleus}\label{yield}
\begin{figure}
    \begin{center}
      \resizebox{0.45\textwidth}{!}{
        \includegraphics{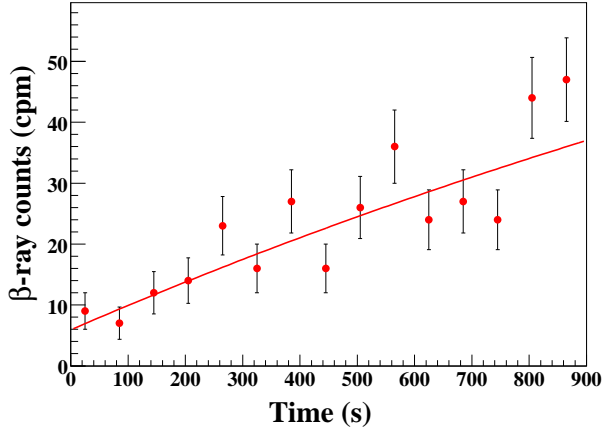}
      }
        \caption{Typical growth curve of $^{199}$Pt nucleus, which was
          measured for the evaluation of relative extraction yield
          $I_{\rm 0}$of $^{199}$Pt nucleus extracted from the KISS.
        }\label{199pt_growth}
    \end{center}
\end{figure}

In the present experiment, the relative extracted yield of $^{199}$Pt
nucleus was measured as a function of wavelength of $\lambda_1$, and
the hyperfine splitting was studied to evaluate the magnetic dipole moment.
Relative extraction yield $I_{\rm 0}$ was deduced from the growth curve during 15
min. irradiation as shown in Fig. \ref{199pt_growth}.
Solid line in Fig. \ref{199pt_growth} is the best fit result, and the
fitting function consists of single exponential function (free
parameter of amplitude $I_{\rm 0}$) and constant background.
The fluctuations of the primary beam intensity, laser powers,
wavelength $\lambda_1$ within 0.2 pm, spacial and temporal overlapping
influenced the number of measured $\beta$-rays emitted from $^{199}$Pt
nucleus. 
The growth curve in Fig. \ref{199pt_growth} reflected the
fluctuations, and, therefore,  the evaluated $I_{\rm 0}$ and its error
$\delta I_{\rm 0}$ take the fluctuations into account automatically.
The $I_{\rm 0}$ and $\delta I_{\rm 0}$ were normalized by the measured
primary beam dose during each measurement.

\subsection{Data analysis of hyperfine splitting}
\begin{figure}
    \begin{center}
      \resizebox{0.45\textwidth}{!}{
        \includegraphics{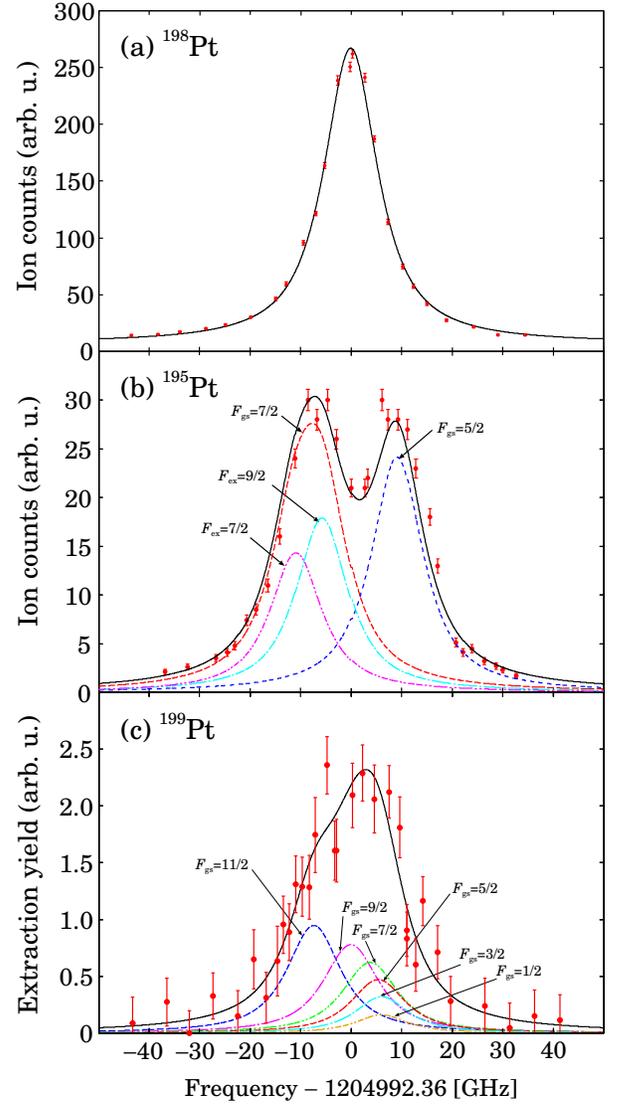}
      }
        \caption{Measured hyperfine splittings of $^{198, 195, 199}$Pt
          nuclei extracted from the KISS. The experimental conditions
          of gas pressure and laser power of $\lambda_1$ were 88 kPa
          and 70 $\mu$J/pulse, respectively. (a) the analysis of the
          single resonance peak of $^{198}$Pt nucleus provides the
          line shape measured under the experimental conditions. (b)
          in the spectrum of $^{195}$Pt nucleus, the line shapes
          labeled as $F_{\rm gs}=$ 5/2 and 7/2 indicate the
          transitions from both the levels to the excited states.
          The line shapes labeled as $F_{\rm ex}=$ 7/2 and 9/2
          indicate the components of the transitions from $F_{\rm
            gs}=$ 7/2 to each excited hyperfine-level.
          (c) in the spectrum of $^{199}$Pt nucleus, the line shapes
          labeled as $F_{\rm gs}=$ 1/2, 3/2, 5/2, 7/2, 9/2, and 11/2
          indicate the transitions from each hyperfine-level on the
          ground state to the excited states as shown in
          Fig. \ref{HFS}--(c), and consist of the transitions from the each
          level to the excited hyperfine-levels in the same manner as
          the transitions of $^{195}$Pt isotope.
        }\label{Pt_HFS}
    \end{center}
\end{figure}

Figure \ref{Pt_HFS} shows the measured hyperfine splittings of $^{198,
  195, 199}$Pt nuclei extracted from the KISS.
The line shape of a resonance peak was determined from the fit of the
measured single peak of $^{198}$Pt ($I^{\pi}=$0$^+$) as shown in Fig. \ref{Pt_HFS}--(a).
The $A_{\rm ex}$ factor for the excited state 5d$^8$6s6p$^5$F$^0_4$
was deduced for the first time from the spectrum analysis of
$^{195}$Pt ($I^{\pi}=$1/2$^-$ and $\mu=$ $+$0.60949(6) $\mu_{\rm N}$
\cite{HIL92}) in Fig. \ref{Pt_HFS}--(b).
In order to determine magnetic dipole moment $\mu$ and IS $\delta \nu$,
the deduced line shape and $A_{\rm ex}$ factor were applied for the spectrum
analysis of $^{199}$Pt ($I^{\pi}=$5/2$^-$) in Fig. \ref{Pt_HFS}--(c).
The details of the analysis are discussed at the following subsections.

\subsubsection{Line shape}
There is no hyperfine splitting for $^{198}$Pt with $I^{\pi}=$ 0$^+$
as shown in Fig. \ref{HFS}--(a), and, therefore, a single resonance peak
was observed as shown in Fig. \ref{Pt_HFS}--(a). 
The line shape is Voigt function which is composed of the convolution of
Gaussian and Lorentzian components. 
The line shape, as shown in Fig. \ref{Pt_HFS}--(a), is governed
mainly by Lorentzian components in the present experiment.
The two Gaussian components are Doppler broadening governed by the
argon gas temperature and laser linewidth, whose width in FWHM were
evaluated to be 1.0 GHz at 300 K and 3.4 GHz, respectively. 
As a result, total Gaussian width of 3.5 GHz was obtained and used as
a fixed parameter in all fits.
Lorentzian components consist of gas-pressure broadening,
laser-power broadening, and the natural width 63 MHz of the present
transition.
The width of Lorentzian components was free parameter in the fitting,
and was deduced to be 11.8(4) GHz. 
As a result, total FWHM 12.8(4) GHz was obtained.
The measured line shape of $^{198}$Pt at the on-line experiment
was applied to other isotopes as the fixed parameter for the fitting of
the spectra.
All fits in the present analysis were performed by using the MINUIT
code \cite{MIN94}.

\subsubsection{Hyperfine coupling constant $A_{\rm ex}$}
The hyperfine splitting of $^{195}$Pt with $I^{\pi}=$ 1/2$^-$ and
$\mu=$ $+$0.60949(6) $\mu_{\rm N}$ \cite{TOI} is governed by
$A_{\rm gs}$ and $A_{\rm ex}$ factors, and there are three transitions
as shown in Fig. \ref{HFS}--(b).
$A_{\rm gs}$ was precisely measured to be 5.70264(23) GHz \cite{BUT84}.
However, $A_{\rm ex}$ of the excited level was unknown. 
The fit to the measured spectrum of $^{195}$Pt in Fig. \ref{Pt_HFS}--(b)
is able to provide the $A_{\rm ex}$ factor.
The line shape of the three transitions are unity, and each resonance
peak position presented by Eq. (\ref{peak-pos}) depends on the $A_{\rm
  ex}$ factor (here, $B_{\rm gs}=B_{\rm ex}=$ 0), which was free as the
fitting parameter.

Another important parameter is an amplitude of each resonance peak.
In order to reduce the fitting parameters and constrain the amplitude,
the relative intensity \cite{KOS11} for each transition was applied
in the present analysis.
The relative intensities of the transitions from the $F_{\rm gs}=$ 5/2
and 7/2 levels are in proportion to the statistical weights (2$F_{\rm gs}+$
1), and are assumed to be 6/14 and 8/14, respectively.
In the present experiment, both $F_{\rm ex}=$ 7/2 and 9/2 levels would be
populated by two transitions from $F_{\rm gs}=$ 7/2 simultaneously due
to the broad line shape.
The relative intensities for the transitions $F_{\rm gs}=$ 7/2
$\rightarrow$ $F_{\rm ex}=$ 7/2 and 9/2 are also in proportion to the
statistical weights (2$F_{\rm ex}+$ 1), and are estimated to be 8/18
and 10/18, respectively.
Finally, the relative intensities for the three transitions of $F_{\rm
  gs}=$ 5/2 $\rightarrow$ $F_{\rm ex}=$ 7/2, $F_{\rm gs}=$ 7/2
$\rightarrow$ $F_{\rm ex}=$ 7/2, and $F_{\rm gs}=$ 7/2 $\rightarrow$
$F_{\rm ex}=$ 9/2 are constrained to be 6/14, 8/14$\times$8/18, and
8/14$\times$10/18, respectively, and one fitting parameter for the
amplitude was used.

The solid black line indicates the best fit to the spectrum in
Fig. \ref{Pt_HFS}, taking into account the isotope shift as the fixed
parameter. 
The IS value was evaluated from the measured ones by using
$^{192,194,196,198}$Pt isotopes, considering the linear relation of
the isotope shift \cite{DUO89,HIL92}.
The $A_{\rm ex}=$ 1.15 $\pm$ 0.11 GHz was obtained as a best fit result.
Even though the experimental conditions such as gas pressure and laser
power of $\lambda_1$ were different, the deduced $A_{\rm ex}$ factor
was consistent with each other.
The $A_{\rm ex}$ factor was used for the spectrum analysis of
$^{199}$Pt isotope.

\subsubsection{Magnetic dipole moment of $^{199}$Pt}
Fig. \ref{Pt_HFS}--(c) shows the measured hyperfine structure of $^{199}$Pt.
Relative extraction yields as mentioned in Sec. \ref{yield} were
measured as a function of $\lambda_1$.
In order to confirm the consistency of the yield $I_{\rm 0}$ during the long-run
measurement, the three-times frequency-scanning in the range $-40\leq
\Delta \nu_{\rm 0} \leq +$40 GHz were performed.
Finally, the consistency of the measured yield $I_{\rm 0}$ was confirmed.

The hyperfine splitting of $F_{\rm gs}=$ 1/2--11/2 and $F_{\rm  ex}=$
3/2--13/2 for $^{199}$Pt with $I=$ 5/2 is governed by $A$ and $B$
factors of the ground and excited states, and there are 15 transitions
as shown in Fig. \ref{HFS}.
Each resonance peak position presented by Eq. (\ref{peak-pos}) depends
on unknown $\mu$, $Q$, $B_{\rm ex}$ factor, and $\Delta \nu_{\rm 0}$
(shift from the center of the gravity $\nu_0=$ 1204992.36 GHz) , which
were free as the fitting parameters.
Here, the $B_{\rm gs}$ factor of the ground state is related to the
$Q$ by $B_{\rm gs}=-Q/0.685$ GHz from Ref. \cite{HIL92}, and this equation
was used in the fitting.
The line shape for each transition was fixed to that evaluated from $^{198}$Pt,
and the relative intensities of the 15 transitions were constrained by
using the statistical weights of (2$F_{\rm gs}+$ 1) and (2$F_{\rm ex}+$ 1)
in the same manner as the fitting procedure of $^{195}$Pt.

The solid black line indicates the best fit to the spectrum in Fig. \ref{Pt_HFS}.
Measured $\mu$, $Q$ and $B_{\rm ex}$ were 0.63 $\pm$ 0.13(stat.) $\pm$
0.03(syst.) $\mu_{\rm N}$, $-$2.7 $\pm$ 3.0 b, $-$1.8 $\pm$ 9.4 GHz, respectively.
Here, the fitting errors were corrected by the square root of reduced
$\chi^2$ 1.14 of the fitting. 
The systematic error stems from the ambiguity of the width of the line
shape and $A_{\rm ex}$.
The width of the spectrum is mainly governed by the magnetic dipole
moment of $^{199}$Pt nucleus. Therefore, it was possible to deduce the
$\mu$ value. 
The $Q$ and $B_{\rm ex}$ were insensitive to the spectrum fitting.
Even though the $Q$ and $B_{\rm ex}$ were artificially assumed
to be 0, this affects the deviation $\Delta \mu=$ 0.015 $\mu_{\rm N}$
which is negligible in considering the present error.

Table. \ref{gfac_table} shows the systematics of magnetic dipole
moments of nuclei with $I^{\pi}=$ 5/2$^-$.
$I^{\pi}=$ 5/2$^-$ is determined by one valence neutron in 2$f_{5/2}$
orbit, whose Schmidt value is evaluated to $+$1.366 $\mu_{\rm N}$.
Although the present $\mu=$ 0.63(13) is smaller than the systematics
of about $+$0.85 for the nuclei, it is consistent with the systematics
by taking the error into account.
The $\mu$ values in Table. \ref{gfac_table} is about 60\% of Schmidt value.
This indicates that large configuration mixing exists in the
wave-function of these states.

\begin{table*}
\begin{center}
  \caption{Systematics of magnetic dipole moments of nuclei with $I^{\pi}=$ 5/2$^-$.
  }\label{gfac_table}
  \begin{tabular}{cccccccc} \hline \hline
Nuclide & $I^{\pi}$ & $t_{\rm 1/2}$ & $E_{\rm x}$ (keV) & $\mu$
($\mu_{\rm N}$)  & $Q$ (b) & Method & Ref. \\ \hline
$^{199}_{78}$Pt & 5/2$^-$ & 30.8(2) min. & 0 &
$+$0.63$\pm$0.13(stat.)$\pm$0.03(syst.)& -2.7$\pm$3.0 & IGLIS & Present work \\
$^{195}_{78}$Pt & 5/2$^-$ & 0.67(3) ns & 130 & $+$0.875(100) & - &
M$\ddot{\rm o}$ssbauer & \cite{WOL72} \\
$^{197}_{78}$Pt & 5/2$^-$ & 16.58(17) ns & 53.088 & $+$0.85(3) & - & TDPAC
& \cite{SAX81}\\
$^{197}_{80}$Hg & 5/2$^-$ & 8.066(8) ns & 134 & $+$0.855(15) & $-$0.081(6)
& TDPAC & \cite{KRI77}\\
$^{199}_{80}$Hg & 5/2$^-$ & 2.45(5) ns & 158 & $+$0.880(33)  & $+$0.95(7) &
TDPAC & \cite{KRI77} \\ \hline \hline
  \end{tabular}
\end{center}
\end{table*}

\subsubsection{Isotope shift and deformation parameter of $^{199}$Pt}
Measured isotope shift (IS) between $^{198}$Pt and $^{199}$Pt was $\delta
\nu^{198,199}=$ 0.98 $\pm$ 0.48 GHz.
Table \ref{is_table} shows the summary of the measured isotope shifts
of the wavelength $\lambda_1=$ 248.792 nm, isotopic variation of the
mean-square charge radius from Ref. \cite{HIL92}, and quadrupole
deformation parameters calculated by using the droplet model \cite{WIL83}.
The IS $\delta \nu^{A,A'} = \nu_0^{A'} - \nu_0^A$ between two isotopes
with mass number $A$ and $A'$ is expressed as
\begin{eqnarray}
\delta \nu^{A, A'} &=& (K_{\rm NMS}+K_{\rm SMS}) \times \frac{m_{\rm
    A'}-m_{\rm A}}{m_{\rm A'}m_{\rm A} } \nonumber \\[1mm]
                 & & + F_{\rm 248} \times \delta<r^2>^{A,A'}. \label{is_dr}
\end{eqnarray}
Here, $K_{\rm NMS}$ and $K_{\rm SMS}$ are factors for normal (NMS) and specific
(SMS) mass shift components, respectively.
The term $F_{\rm 248} \times \delta<r^2>^{A,A'}$ is the field shift (FS)
component expressed by using the electronic factor $F_{\rm 248}$ for the
transition $\lambda_1=$ 248.792 nm and the isotopic variation of the
mean-square charge radius $\delta<r^2>^{A,A'}$.

\begin{figure}
    \begin{center}
      \resizebox{0.45\textwidth}{!}{
        \includegraphics{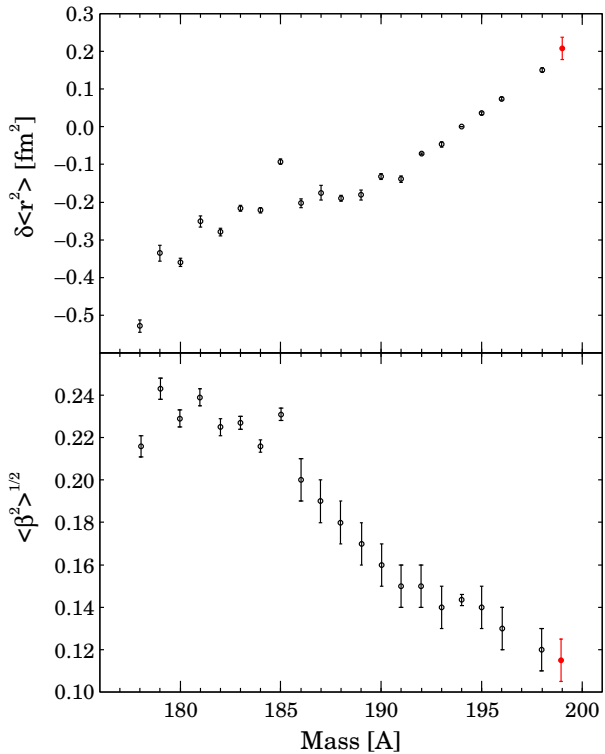}
      }
        \caption{Systematics of mean square charge radius variation
          $\delta<r^2>^{194,A}$ and quadrupole deformation parameter
          $|<\beta_2^2>^{1/2}|$ as a function of mass number $A$ of
          the platinum isotopes. Open and closed circles present the
          values from Ref. \cite{HIL92,BLA99} and by the present work,
          respectively.
        }\label{pt_dr_beta}
    \end{center}
\end{figure}

In order to deduce $\delta<r^2>^{A,A'}$ from measured
$\delta\nu^{A,A'}$, the Eq. (\ref{is_dr}) can be practically modified
as follows;
\begin{eqnarray}
\delta<r^2>^{A,A'}\frac{m_{\rm A'}m_{\rm A}}{m_{\rm A'}-m_{\rm A}} &=&
-\frac{K_{\rm MS}}{F_{248}} \nonumber \\[1mm]
  & & + \frac{1}{F_{248}}\delta \nu^{A, A'}\frac{m_{\rm A'}m_{\rm A}}{m_{\rm A'}-m_{\rm A}}. \label{is_dr_mod}
\end{eqnarray}
Here, $K_{\rm MS} = K_{\rm FMS} + K_{\rm SMS}$.
If both $\delta \nu^{A,A'}$ and $\delta<r^2>^{A,A'}$ for
stable isotopes were known, $K_{\rm MS}$ and $F_{248}$ can be
evaluated for deducing the $\delta<r^2>^{A,A'}$ from measured $\delta \nu^{A,A'}$.
Generally, the term of the mass shift ($\sim$ a few 10 MHz) is about
two order of magnitude smaller than the term of the field shift
($\sim$ a few GHz) in the present high $Z=$ 78 region \cite{BLA13},
and the mass shift term can be negligible in the present case.
The factor $F_{248}$ was evaluated to 17.0 $\pm$ 1.8 GHz/fm$^2$
from the fit using Eq. (\ref{is_dr_mod}) between the measured
$\delta \nu^{A,A'}$ and the reported $\delta <r^2>^{A,A'}$
\cite{HIL92} for $^{\rm even}$Pt in Table \ref{is_table}.
Finally, the $\delta<r^2>^{198,199}=$ 0.057(29) fm$^2$
($\delta<r^2>^{194,199}=$ 0.208(30)) was obtained from the $\delta
\nu^{198,199}=$ 0.98(48) GHz.
Figure \ref{pt_dr_beta} shows the mean square charge radius variation
of the platinum isotopes \cite{HIL92,BLA99}.
The evaluated $\delta<r^2>^{194,199}$ shows the linear relation of
platinum isotopes with 193$\leq A \leq$ 198.

Model-dependent quadrupole deformation parameter $<\beta_2^2>$ can be
extracted from the measured $\delta \nu^{A,A'}$ and $\delta <r^2>^{A,A'}$.
The details of the extraction procedure and the utilization for the
platinum isotopes  were reported in Ref. \cite{AHM85} and
Ref. \cite{HIL92}, respectively.
The evaluated parameters $\delta <\beta_2^2>$ and
$|<\beta_2^2>^{1/2}|$ were listed in Table \ref{is_table}.
The absolute deformation parameter $|<\beta_2^2>^{1/2}|$ for
$^{199}$Pt was evaluated from the relation of $<\beta_2^2>^{A} =
<\beta_2^2>^{194} + \delta <\beta_2^2>^{194,A}$.
Here, the parameter $<\beta_{2}^{2}>^{194}$ was deduced by using
the precisely measured $B$($E$2,2$^+ \rightarrow$ 0$^+$) value of
$^{194}$Pt \cite{RAM89}, and the evaluated parameters
$<\beta_2^2>^{\rm even Pt}$ were in good agreement with the known
$B$($E$2) values \cite{RAM89} for $^{\rm even}$Pt.
The evaluated $\delta <\beta_2^2>$ also shows the linear relation of
platinum isotopes with 193$\leq A \leq$ 198.
The $|<\beta_2^2>^{1/2}|$ for $^{199}$Pt becomes smaller with
increasing neutron number toward $N=$ 126 as shown in
Fig. \ref{pt_dr_beta}, indicating the spherical shape for platinum
isotope with $N=$ 126.

\begin{table}
\begin{center}
  \caption{Measured isotope shifts $\delta \nu^{194,A}_{\rm exp}$ and
    $\delta \nu^{198,199}_{\rm exp}$, isotopic variation of the
    mean-square charge radius $\delta <r^2>^{194,199}$ and
    $\delta <r^2>^{194,A}$ from Ref. \cite{HIL92}. Quadrupole
    deformation parameters $\delta <\beta>^{194,A}$ and
    $<\beta^2_2>^{1/2}$ calculated by using Droplet model \cite{WIL83}.
  }\label{is_table}
  \begin{tabular}{ccccc} \hline \hline
 $A$ & $\delta \nu^{194,A}_{\rm exp}$ & $\delta <r^2>^{194,A}$ &  $\delta <\beta>^{194,A}$ & $|<\beta^2_2>^{1/2}|$ \\ 
  &  [GHz] & [fm$^2$] &  &  \\ \hline
199 & 0.98(48)$^a$ & 0.208(30)$^b$ & $-$0.067(10)$^b$ & 0.115(10)$^b$ \\ 
198 & 2.43(23)$^b$ & 0.151(6) & $-$0.0050(7) & 0.12(1) \\ 
196 & 1.39(27)$^b$ & 0.074(4)  & $-$0.0027(3) & 0.13(1) \\ 
194 & 0.000 & 0.000  & 0.000 & 0.1434(26) \\ 
192 & $-$1.43(27)$^b$ & $-$0.072(3) & 0.0033(9) & 0.15(1) \\ \hline \hline
  \end{tabular}
\end{center}
$^a$~Present work : measured isotope shift $\delta \nu^{198,199}_{\rm exp}$.\\
$^b$~Present work\\
\end{table}

\section{SUMMARY}
Neutron-rich nucleus $^{199}$Pt ($I^{\pi}=$ 5/2$^-$) was produced by
multi-nucleon transfer reaction at the KISS, where the nuclear
spectroscopy around $N=$ 126 have been performed from the aspect of
an astrophysical interest as well as the nuclear structure.
Magnetic dipole moments provide the crucial information about the
nuclear wave-function, which is essential for the understanding of the
competition between allowed Gamow-Teller (GT) and first-forbidden (FF)
beta-decay transitions in the nuclei around $N$ = 126.

By in-gas-cell laser ionization spectroscopy performed at the KISS,
magnetic dipole moment $\mu$ and mean-square charge radius variation
$\delta<r^2>^{194,199}$ of $^{199}$Pt have been determined for the
first time to be $+$0.63(13) $\mu_{\rm  N}$ and 0.208(30) fm$^2$. 
Measured magnetic dipole moment is consistent with the systematics of
those of nuclei with $I^{\pi}=$ 5/2$^-$.
The evaluated parameters $\delta <\beta_2^2>$ and
$|<\beta_2^2>^{1/2}|$ were consistent with the linear relation of platinum isotopes
with 193$\leq A \leq$ 198.
The behavior with increasing neutron number toward $N=$ 126 would
indicate the shape change to spherical shape of platinum isotopes.

By taking the advantage of the MNT reaction combined with the KISS,
systematical nuclear spectroscopy is planed for the neutron-rich
isotopes of the refractory elements such as Pt, Ir, Os, Re, W and Ta.

\begin{acknowledgments}
This experiment was performed at RI Beam Factory operated by RIKEN
Nishina Center and CNS, the University of Tokyo.
The authors acknowledge the staff of the accelerator for their support.
This work has been supported by Grant-in-Aids for
Scientific Research (A) (grant no. 23244060, 26247044 and 15H02096)
and for young scientists (B) (grant no. 24740180) from the Japan
Society for the Promotion of Science (JSPS).
\end{acknowledgments}

\bibliography{199pt}

\providecommand{\noopsort}[1]{}\providecommand{\singleletter}[1]{#1}%
\begin{thebibliography}{40}%
\makeatletter
\providecommand \@ifxundefined [1]{%
 \@ifx{#1\undefined}
}%
\providecommand \@ifnum [1]{%
 \ifnum #1\expandafter \@firstoftwo
 \else \expandafter \@secondoftwo
 \fi
}%
\providecommand \@ifx [1]{%
 \ifx #1\expandafter \@firstoftwo
 \else \expandafter \@secondoftwo
 \fi
}%
\providecommand \natexlab [1]{#1}%
\providecommand \enquote  [1]{``#1''}%
\providecommand \bibnamefont  [1]{#1}%
\providecommand \bibfnamefont [1]{#1}%
\providecommand \citenamefont [1]{#1}%
\providecommand \href@noop [0]{\@secondoftwo}%
\providecommand \href [0]{\begingroup \@sanitize@url \@href}%
\providecommand \@href[1]{\@@startlink{#1}\@@href}%
\providecommand \@@href[1]{\endgroup#1\@@endlink}%
\providecommand \@sanitize@url [0]{\catcode `\\12\catcode `\$12\catcode
  `\&12\catcode `\#12\catcode `\^12\catcode `\_12\catcode `\%12\relax}%
\providecommand \@@startlink[1]{}%
\providecommand \@@endlink[0]{}%
\providecommand \url  [0]{\begingroup\@sanitize@url \@url }%
\providecommand \@url [1]{\endgroup\@href {#1}{\urlprefix }}%
\providecommand \urlprefix  [0]{URL }%
\providecommand \Eprint [0]{\href }%
\providecommand \doibase [0]{http://dx.doi.org/}%
\providecommand \selectlanguage [0]{\@gobble}%
\providecommand \bibinfo  [0]{\@secondoftwo}%
\providecommand \bibfield  [0]{\@secondoftwo}%
\providecommand \translation [1]{[#1]}%
\providecommand \BibitemOpen [0]{}%
\providecommand \bibitemStop [0]{}%
\providecommand \bibitemNoStop [0]{.\EOS\space}%
\providecommand \EOS [0]{\spacefactor3000\relax}%
\providecommand \BibitemShut  [1]{\csname bibitem#1\endcsname}%
\let\auto@bib@innerbib\@empty
%</preamble>
\bibitem [{\citenamefont {Borzov}(2003)}]{BOR03}%
  \BibitemOpen
  \bibfield  {author} {\bibinfo {author} {\bibfnamefont {I.~N.}\ \bibnamefont
  {Borzov}},\ }\href@noop {} {\bibfield  {journal} {\bibinfo  {journal} {Phys.\
  Rev.\ C}\ }\textbf {\bibinfo {volume} {67}},\ \bibinfo {pages} {025802}
  (\bibinfo {year} {2003})}\BibitemShut {NoStop}%
\bibitem [{\citenamefont {M$\ddot{\rm o}$ller}\ \emph
  {et~al.}(2003)\citenamefont {M$\ddot{\rm o}$ller}, \citenamefont {Pfeiffer},\
  and\ \citenamefont {Kratz}}]{MOL03}%
  \BibitemOpen
  \bibfield  {author} {\bibinfo {author} {\bibfnamefont {P.}~\bibnamefont
  {M$\ddot{\rm o}$ller}}, \bibinfo {author} {\bibfnamefont {B.}~\bibnamefont
  {Pfeiffer}}, \ and\ \bibinfo {author} {\bibfnamefont {K.-L.}\ \bibnamefont
  {Kratz}},\ }\href@noop {} {\bibfield  {journal} {\bibinfo  {journal} {Phys.\
  Rev.\ C}\ }\textbf {\bibinfo {volume} {67}},\ \bibinfo {pages} {055802}
  (\bibinfo {year} {2003})}\BibitemShut {NoStop}%
\bibitem [{\citenamefont {Koura}\ \emph {et~al.}(2005)\citenamefont {Koura},
  \citenamefont {Tachibana}, \citenamefont {Uno},\ and\ \citenamefont
  {Yamada}}]{KOU05}%
  \BibitemOpen
  \bibfield  {author} {\bibinfo {author} {\bibfnamefont {H.}~\bibnamefont
  {Koura}}, \bibinfo {author} {\bibfnamefont {T.}~\bibnamefont {Tachibana}},
  \bibinfo {author} {\bibfnamefont {M.}~\bibnamefont {Uno}}, \ and\ \bibinfo
  {author} {\bibfnamefont {M.}~\bibnamefont {Yamada}},\ }\href@noop {}
  {\bibfield  {journal} {\bibinfo  {journal} {Prog.\ Theor.\ Phys.}\ }\textbf
  {\bibinfo {volume} {113}},\ \bibinfo {pages} {305} (\bibinfo {year}
  {2005})}\BibitemShut {NoStop}%
\bibitem [{\citenamefont {M$\ddot{\rm o}$ller}\ \emph
  {et~al.}(1997)\citenamefont {M$\ddot{\rm o}$ller}, \citenamefont {Nix},\ and\
  \citenamefont {Kratz}}]{MOL97}%
  \BibitemOpen
  \bibfield  {author} {\bibinfo {author} {\bibfnamefont {P.}~\bibnamefont
  {M$\ddot{\rm o}$ller}}, \bibinfo {author} {\bibfnamefont {J.~R.}\
  \bibnamefont {Nix}}, \ and\ \bibinfo {author} {\bibfnamefont {K.-L.}\
  \bibnamefont {Kratz}},\ }\href@noop {} {\bibfield  {journal} {\bibinfo
  {journal} {At.\ Data\ Nucl.\ Data Tables}\ }\textbf {\bibinfo {volume}
  {66}},\ \bibinfo {pages} {131} (\bibinfo {year} {1997})}\BibitemShut
  {NoStop}%
\bibitem [{\citenamefont {Engel}\ \emph {et~al.}(1999)\citenamefont {Engel},
  \citenamefont {Bender}, \citenamefont {Dobaczewski}, \citenamefont
  {Nazarewicz},\ and\ \citenamefont {Surman}}]{ENG99}%
  \BibitemOpen
  \bibfield  {author} {\bibinfo {author} {\bibfnamefont {J.}~\bibnamefont
  {Engel}}, \bibinfo {author} {\bibfnamefont {M.}~\bibnamefont {Bender}},
  \bibinfo {author} {\bibfnamefont {J.}~\bibnamefont {Dobaczewski}}, \bibinfo
  {author} {\bibfnamefont {W.}~\bibnamefont {Nazarewicz}}, \ and\ \bibinfo
  {author} {\bibfnamefont {R.}~\bibnamefont {Surman}},\ }\href@noop {}
  {\bibfield  {journal} {\bibinfo  {journal} {Phys.\ Rev.\ C}\ }\textbf
  {\bibinfo {volume} {60}},\ \bibinfo {pages} {014302} (\bibinfo {year}
  {1999})}\BibitemShut {NoStop}%
\bibitem [{\citenamefont {Martinez-Pinedo}(2001)}]{PIN01}%
  \BibitemOpen
  \bibfield  {author} {\bibinfo {author} {\bibfnamefont {G.}~\bibnamefont
  {Martinez-Pinedo}},\ }\href@noop {} {\bibfield  {journal} {\bibinfo
  {journal} {Nucl.\ Phys.\ A}\ }\textbf {\bibinfo {volume} {688}},\ \bibinfo
  {pages} {357c} (\bibinfo {year} {2001})}\BibitemShut {NoStop}%
\bibitem [{\citenamefont {Langanke}\ and\ \citenamefont
  {Martinez-Pinedo}(2003)}]{LAN03}%
  \BibitemOpen
  \bibfield  {author} {\bibinfo {author} {\bibfnamefont {K.}~\bibnamefont
  {Langanke}}\ and\ \bibinfo {author} {\bibfnamefont {G.}~\bibnamefont
  {Martinez-Pinedo}},\ }\href@noop {} {\bibfield  {journal} {\bibinfo
  {journal} {Rev.\ Mod.\ Phys.}\ }\textbf {\bibinfo {volume} {75}},\ \bibinfo
  {pages} {819} (\bibinfo {year} {2003})}\BibitemShut {NoStop}%
\bibitem [{\citenamefont {KUTY}()}]{KUTY}%
  \BibitemOpen
  \bibfield  {author} {\bibinfo {author} {\bibnamefont {KUTY}},\ }\href@noop {}
  {\bibinfo  {journal}
  {http://wwwndc.jaea.go.jp/nucldata/beta-decay-properties.pdf}\ }\BibitemShut
  {NoStop}%
\bibitem [{\citenamefont {Burbidge}\ \emph {et~al.}(1957)\citenamefont
  {Burbidge}, \citenamefont {Burbidge}, \citenamefont {Fowler},\ and\
  \citenamefont {Hoyle}}]{BUR57}%
  \BibitemOpen
\bibfield  {journal} {  }\bibfield  {author} {\bibinfo {author} {\bibfnamefont
  {E.~M.}\ \bibnamefont {Burbidge}}, \bibinfo {author} {\bibfnamefont {G.~R.}\
  \bibnamefont {Burbidge}}, \bibinfo {author} {\bibfnamefont {W.~A.}\
  \bibnamefont {Fowler}}, \ and\ \bibinfo {author} {\bibfnamefont
  {F.}~\bibnamefont {Hoyle}},\ }\href@noop {} {\bibfield  {journal} {\bibinfo
  {journal} {Rev.\ Mod.\ Phys.}\ }\textbf {\bibinfo {volume} {29}},\ \bibinfo
  {pages} {547} (\bibinfo {year} {1957})}\BibitemShut {NoStop}%
\bibitem [{\citenamefont {Mumpower}\ \emph {et~al.}(2016)\citenamefont
  {Mumpower}, \citenamefont {Surman}, \citenamefont {McLaughlin},\ and\
  \citenamefont {Aprahamian}}]{MUM16}%
  \BibitemOpen
  \bibfield  {author} {\bibinfo {author} {\bibfnamefont {M.~R.}\ \bibnamefont
  {Mumpower}}, \bibinfo {author} {\bibfnamefont {R.}~\bibnamefont {Surman}},
  \bibinfo {author} {\bibfnamefont {G.~C.}\ \bibnamefont {McLaughlin}}, \ and\
  \bibinfo {author} {\bibfnamefont {A.}~\bibnamefont {Aprahamian}},\
  }\href@noop {} {\bibfield  {journal} {\bibinfo  {journal} {Prog.\ Part.\
  Nucl.\ Phys.}\ }\textbf {\bibinfo {volume} {86}},\ \bibinfo {pages} {86}
  (\bibinfo {year} {2016})}\BibitemShut {NoStop}%
\bibitem [{\citenamefont {Stone}(2011)}]{STO11}%
  \BibitemOpen
  \bibfield  {author} {\bibinfo {author} {\bibfnamefont {N.~J.}\ \bibnamefont
  {Stone}},\ }\href@noop {} {\emph {\bibinfo {title} {Table of Nuclear Magnetic
  Dipole and Electric Quadrupole Moments}}}\ (\bibinfo {year}
  {2011})\BibitemShut {NoStop}%
\bibitem [{\citenamefont {Jeong}\ \emph {et~al.}(2010)\citenamefont {Jeong},
  \citenamefont {Imai}, \citenamefont {Ishiyama}, \citenamefont {Hirayama},
  \citenamefont {Miyatake},\ and\ \citenamefont {Watanabe}}]{JEO10}%
  \BibitemOpen
  \bibfield  {author} {\bibinfo {author} {\bibfnamefont {S.~C.}\ \bibnamefont
  {Jeong}}, \bibinfo {author} {\bibfnamefont {N.}~\bibnamefont {Imai}},
  \bibinfo {author} {\bibfnamefont {H.}~\bibnamefont {Ishiyama}}, \bibinfo
  {author} {\bibfnamefont {Y.}~\bibnamefont {Hirayama}}, \bibinfo {author}
  {\bibfnamefont {H.}~\bibnamefont {Miyatake}}, \ and\ \bibinfo {author}
  {\bibfnamefont {Y.~X.}\ \bibnamefont {Watanabe}},\ }\href@noop {} {\bibfield
  {journal} {\bibinfo  {journal} {KEK Report}\ }\textbf {\bibinfo {volume}
  {2010-2}} (\bibinfo {year} {2010})}\BibitemShut {NoStop}%
\bibitem [{\citenamefont {Dasso}\ \emph {et~al.}(1994)\citenamefont {Dasso},
  \citenamefont {Pollarolo},\ and\ \citenamefont {Winther}}]{DAS94}%
  \BibitemOpen
  \bibfield  {author} {\bibinfo {author} {\bibfnamefont {C.~H.}\ \bibnamefont
  {Dasso}}, \bibinfo {author} {\bibfnamefont {G.}~\bibnamefont {Pollarolo}}, \
  and\ \bibinfo {author} {\bibfnamefont {A.}~\bibnamefont {Winther}},\
  }\href@noop {} {\bibfield  {journal} {\bibinfo  {journal} {Phys.\ Rev.\
  Lett.}\ }\textbf {\bibinfo {volume} {73}},\ \bibinfo {pages} {1907} (\bibinfo
  {year} {1994})}\BibitemShut {NoStop}%
\bibitem [{\citenamefont {Watanabe}\ \emph {et~al.}(2015)\citenamefont
  {Watanabe}, \citenamefont {Kim}, \citenamefont {Jeong}, \citenamefont
  {Hirayama}, \citenamefont {Imai}, \citenamefont {Ishiyama}, \citenamefont
  {Jung}, \citenamefont {Miyatake}, \citenamefont {Choi}, \citenamefont {Song},
  \citenamefont {Clement}, \citenamefont {de~France}, \citenamefont {Navin},
  \citenamefont {Rejmund}, \citenamefont {Schmitt}, \citenamefont {Pollarolo},
  \citenamefont {Corradi}, \citenamefont {Fioretto}, \citenamefont {Montanari},
  \citenamefont {Niikura}, \citenamefont {Suzuki}, \citenamefont {Nishibata},\
  and\ \citenamefont {Takatsu}}]{WAT15}%
  \BibitemOpen
  \bibfield  {author} {\bibinfo {author} {\bibfnamefont {Y.~X.}\ \bibnamefont
  {Watanabe}}, \bibinfo {author} {\bibfnamefont {Y.~H.}\ \bibnamefont {Kim}},
  \bibinfo {author} {\bibfnamefont {S.~C.}\ \bibnamefont {Jeong}}, \bibinfo
  {author} {\bibfnamefont {Y.}~\bibnamefont {Hirayama}}, \bibinfo {author}
  {\bibfnamefont {N.}~\bibnamefont {Imai}}, \bibinfo {author} {\bibfnamefont
  {H.}~\bibnamefont {Ishiyama}}, \bibinfo {author} {\bibfnamefont {H.~S.}\
  \bibnamefont {Jung}}, \bibinfo {author} {\bibfnamefont {H.}~\bibnamefont
  {Miyatake}}, \bibinfo {author} {\bibfnamefont {S.}~\bibnamefont {Choi}},
  \bibinfo {author} {\bibfnamefont {J.~S.}\ \bibnamefont {Song}}, \bibinfo
  {author} {\bibfnamefont {E.}~\bibnamefont {Clement}}, \bibinfo {author}
  {\bibfnamefont {G.}~\bibnamefont {de~France}}, \bibinfo {author}
  {\bibfnamefont {A.}~\bibnamefont {Navin}}, \bibinfo {author} {\bibfnamefont
  {M.}~\bibnamefont {Rejmund}}, \bibinfo {author} {\bibfnamefont
  {C.}~\bibnamefont {Schmitt}}, \bibinfo {author} {\bibfnamefont
  {G.}~\bibnamefont {Pollarolo}}, \bibinfo {author} {\bibfnamefont
  {L.}~\bibnamefont {Corradi}}, \bibinfo {author} {\bibfnamefont
  {E.}~\bibnamefont {Fioretto}}, \bibinfo {author} {\bibfnamefont
  {D.}~\bibnamefont {Montanari}}, \bibinfo {author} {\bibfnamefont
  {M.}~\bibnamefont {Niikura}}, \bibinfo {author} {\bibfnamefont
  {D.}~\bibnamefont {Suzuki}}, \bibinfo {author} {\bibfnamefont
  {H.}~\bibnamefont {Nishibata}}, \ and\ \bibinfo {author} {\bibfnamefont
  {J.}~\bibnamefont {Takatsu}},\ }\href@noop {} {\bibfield  {journal} {\bibinfo
   {journal} {Phys.\ Rev.\ Lett.}\ }\textbf {\bibinfo {volume} {115}},\
  \bibinfo {pages} {172503} (\bibinfo {year} {2015})}\BibitemShut {NoStop}%
\bibitem [{\citenamefont {Hirayama}\ \emph {et~al.}(2015)\citenamefont
  {Hirayama}, \citenamefont {Watanabe}, \citenamefont {Imai}, \citenamefont
  {Ishiyama}, \citenamefont {Jeong}, \citenamefont {Miyatake}, \citenamefont
  {Oyaizu}, \citenamefont {Kimura}, \citenamefont {Mukai}, \citenamefont {Kim},
  \citenamefont {Sonoda}, \citenamefont {Wada}, \citenamefont {Huyse},
  \citenamefont {Kudryavtsev},\ and\ \citenamefont {Duppen}}]{HIR15}%
  \BibitemOpen
  \bibfield  {author} {\bibinfo {author} {\bibfnamefont {Y.}~\bibnamefont
  {Hirayama}}, \bibinfo {author} {\bibfnamefont {Y.~X.}\ \bibnamefont
  {Watanabe}}, \bibinfo {author} {\bibfnamefont {N.}~\bibnamefont {Imai}},
  \bibinfo {author} {\bibfnamefont {H.}~\bibnamefont {Ishiyama}}, \bibinfo
  {author} {\bibfnamefont {S.~C.}\ \bibnamefont {Jeong}}, \bibinfo {author}
  {\bibfnamefont {H.}~\bibnamefont {Miyatake}}, \bibinfo {author}
  {\bibfnamefont {M.}~\bibnamefont {Oyaizu}}, \bibinfo {author} {\bibfnamefont
  {S.}~\bibnamefont {Kimura}}, \bibinfo {author} {\bibfnamefont
  {M.}~\bibnamefont {Mukai}}, \bibinfo {author} {\bibfnamefont {Y.~H.}\
  \bibnamefont {Kim}}, \bibinfo {author} {\bibfnamefont {T.}~\bibnamefont
  {Sonoda}}, \bibinfo {author} {\bibfnamefont {M.}~\bibnamefont {Wada}},
  \bibinfo {author} {\bibfnamefont {M.}~\bibnamefont {Huyse}}, \bibinfo
  {author} {\bibfnamefont {Y.}~\bibnamefont {Kudryavtsev}}, \ and\ \bibinfo
  {author} {\bibfnamefont {P.~V.}\ \bibnamefont {Duppen}},\ }\href@noop {}
  {\bibfield  {journal} {\bibinfo  {journal} {Nucl.\ Instrum.\ Methods\ Phys.\
  Res.\ B}\ }\textbf {\bibinfo {volume} {353}},\ \bibinfo {pages} {4} (\bibinfo
  {year} {2015})}\BibitemShut {NoStop}%
\bibitem [{\citenamefont {Hirayama}\ \emph {et~al.}(2016)\citenamefont
  {Hirayama}, \citenamefont {Watanabe}, \citenamefont {Imai}, \citenamefont
  {Ishiyama}, \citenamefont {Jeong}, \citenamefont {Jung}, \citenamefont
  {Miyatake}, \citenamefont {Oyaizu}, \citenamefont {Kimura}, \citenamefont
  {Mukai}, \citenamefont {Kim}, \citenamefont {Sonoda}, \citenamefont {Wada},
  \citenamefont {Huyse}, \citenamefont {Kudryavtsev},\ and\ \citenamefont
  {Duppen}}]{HIR16}%
  \BibitemOpen
  \bibfield  {author} {\bibinfo {author} {\bibfnamefont {Y.}~\bibnamefont
  {Hirayama}}, \bibinfo {author} {\bibfnamefont {Y.~X.}\ \bibnamefont
  {Watanabe}}, \bibinfo {author} {\bibfnamefont {N.}~\bibnamefont {Imai}},
  \bibinfo {author} {\bibfnamefont {H.}~\bibnamefont {Ishiyama}}, \bibinfo
  {author} {\bibfnamefont {S.~C.}\ \bibnamefont {Jeong}}, \bibinfo {author}
  {\bibfnamefont {H.~S.}\ \bibnamefont {Jung}}, \bibinfo {author}
  {\bibfnamefont {H.}~\bibnamefont {Miyatake}}, \bibinfo {author}
  {\bibfnamefont {M.}~\bibnamefont {Oyaizu}}, \bibinfo {author} {\bibfnamefont
  {S.}~\bibnamefont {Kimura}}, \bibinfo {author} {\bibfnamefont
  {M.}~\bibnamefont {Mukai}}, \bibinfo {author} {\bibfnamefont {Y.~H.}\
  \bibnamefont {Kim}}, \bibinfo {author} {\bibfnamefont {T.}~\bibnamefont
  {Sonoda}}, \bibinfo {author} {\bibfnamefont {M.}~\bibnamefont {Wada}},
  \bibinfo {author} {\bibfnamefont {M.}~\bibnamefont {Huyse}}, \bibinfo
  {author} {\bibfnamefont {Y.}~\bibnamefont {Kudryavtsev}}, \ and\ \bibinfo
  {author} {\bibfnamefont {P.~V.}\ \bibnamefont {Duppen}},\ }\href@noop {}
  {\bibfield  {journal} {\bibinfo  {journal} {Nucl.\ Instrum.\ Methods\ Phys.\
  Res.\ B}\ }\textbf {\bibinfo {volume} {376}},\ \bibinfo {pages} {52}
  (\bibinfo {year} {2016})}\BibitemShut {NoStop}%
\bibitem [{\citenamefont {Cocolios}\ \emph {et~al.}(2010)\citenamefont
  {Cocolios}, \citenamefont {Andreyev}, \citenamefont {Bastin}, \citenamefont
  {Bree}, \citenamefont {B$\ddot{\rm u}$scher}, \citenamefont {Elseviers},
  \citenamefont {Gentens}, \citenamefont {Huyse}, \citenamefont {Kudryavtsev},
  \citenamefont {Pauwels}, \citenamefont {Sonoda}, \citenamefont {den Bergh},\
  and\ \citenamefont {Duppen}}]{COC10}%
  \BibitemOpen
  \bibfield  {author} {\bibinfo {author} {\bibfnamefont {T.~E.}\ \bibnamefont
  {Cocolios}}, \bibinfo {author} {\bibfnamefont {A.~N.}\ \bibnamefont
  {Andreyev}}, \bibinfo {author} {\bibfnamefont {B.}~\bibnamefont {Bastin}},
  \bibinfo {author} {\bibfnamefont {N.}~\bibnamefont {Bree}}, \bibinfo {author}
  {\bibfnamefont {J.}~\bibnamefont {B$\ddot{\rm u}$scher}}, \bibinfo {author}
  {\bibfnamefont {J.}~\bibnamefont {Elseviers}}, \bibinfo {author}
  {\bibfnamefont {J.}~\bibnamefont {Gentens}}, \bibinfo {author} {\bibfnamefont
  {M.}~\bibnamefont {Huyse}}, \bibinfo {author} {\bibfnamefont
  {Y.}~\bibnamefont {Kudryavtsev}}, \bibinfo {author} {\bibfnamefont
  {D.}~\bibnamefont {Pauwels}}, \bibinfo {author} {\bibfnamefont
  {T.}~\bibnamefont {Sonoda}}, \bibinfo {author} {\bibfnamefont {P.~V.}\
  \bibnamefont {den Bergh}}, \ and\ \bibinfo {author} {\bibfnamefont {P.~V.}\
  \bibnamefont {Duppen}},\ }\href@noop {} {\bibfield  {journal} {\bibinfo
  {journal} {Phys.\ Rev.\ C}\ }\textbf {\bibinfo {volume} {81}},\ \bibinfo
  {pages} {014314} (\bibinfo {year} {2010})}\BibitemShut {NoStop}%
\bibitem [{\citenamefont {Ferrer}\ \emph {et~al.}(2014)\citenamefont {Ferrer},
  \citenamefont {Bree}, \citenamefont {Cocolios}, \citenamefont {Darby},
  \citenamefont {Witte}, \citenamefont {Dexters}, \citenamefont {Diriken},
  \citenamefont {Elseviers}, \citenamefont {Franchoo}, \citenamefont {Huyse},
  \citenamefont {Kesteloot}, \citenamefont {Kudryavtsev}, \citenamefont
  {Pauwels}, \citenamefont {Radulov}, \citenamefont {Roger}, \citenamefont
  {Savajols}, \citenamefont {Duppen},\ and\ \citenamefont {Venhart}}]{FER14}%
  \BibitemOpen
  \bibfield  {author} {\bibinfo {author} {\bibfnamefont {R.}~\bibnamefont
  {Ferrer}}, \bibinfo {author} {\bibfnamefont {N.}~\bibnamefont {Bree}},
  \bibinfo {author} {\bibfnamefont {T.~E.}\ \bibnamefont {Cocolios}}, \bibinfo
  {author} {\bibfnamefont {I.~G.}\ \bibnamefont {Darby}}, \bibinfo {author}
  {\bibfnamefont {H.~D.}\ \bibnamefont {Witte}}, \bibinfo {author}
  {\bibfnamefont {W.}~\bibnamefont {Dexters}}, \bibinfo {author} {\bibfnamefont
  {J.}~\bibnamefont {Diriken}}, \bibinfo {author} {\bibfnamefont
  {J.}~\bibnamefont {Elseviers}}, \bibinfo {author} {\bibfnamefont
  {S.}~\bibnamefont {Franchoo}}, \bibinfo {author} {\bibfnamefont
  {M.}~\bibnamefont {Huyse}}, \bibinfo {author} {\bibfnamefont
  {N.}~\bibnamefont {Kesteloot}}, \bibinfo {author} {\bibfnamefont
  {Y.}~\bibnamefont {Kudryavtsev}}, \bibinfo {author} {\bibfnamefont
  {D.}~\bibnamefont {Pauwels}}, \bibinfo {author} {\bibfnamefont
  {D.}~\bibnamefont {Radulov}}, \bibinfo {author} {\bibfnamefont
  {T.}~\bibnamefont {Roger}}, \bibinfo {author} {\bibfnamefont
  {H.}~\bibnamefont {Savajols}}, \bibinfo {author} {\bibfnamefont {P.~V.}\
  \bibnamefont {Duppen}}, \ and\ \bibinfo {author} {\bibfnamefont
  {M.}~\bibnamefont {Venhart}},\ }\href@noop {} {\bibfield  {journal} {\bibinfo
   {journal} {Phys.\ Lett.\ B}\ }\textbf {\bibinfo {volume} {728}},\ \bibinfo
  {pages} {191} (\bibinfo {year} {2014})}\BibitemShut {NoStop}%
\bibitem [{\citenamefont {Kudryavtsev}\ \emph {et~al.}(2013)\citenamefont
  {Kudryavtsev}, \citenamefont {Ferrer}, \citenamefont {Huyse}, \citenamefont
  {den Bergh},\ and\ \citenamefont {Duppen}}]{KUD13}%
  \BibitemOpen
  \bibfield  {author} {\bibinfo {author} {\bibfnamefont {Y.}~\bibnamefont
  {Kudryavtsev}}, \bibinfo {author} {\bibfnamefont {R.}~\bibnamefont {Ferrer}},
  \bibinfo {author} {\bibfnamefont {M.}~\bibnamefont {Huyse}}, \bibinfo
  {author} {\bibfnamefont {P.~V.}\ \bibnamefont {den Bergh}}, \ and\ \bibinfo
  {author} {\bibfnamefont {P.~V.}\ \bibnamefont {Duppen}},\ }\href@noop {}
  {\bibfield  {journal} {\bibinfo  {journal} {Nucl.\ Instrum.\ Methods\ Phys.\
  Res.\ B}\ }\textbf {\bibinfo {volume} {297}},\ \bibinfo {pages} {7} (\bibinfo
  {year} {2013})}\BibitemShut {NoStop}%
\bibitem [{\citenamefont {Hirayama}\ \emph {et~al.}(2014)\citenamefont
  {Hirayama}, \citenamefont {Mukai}, \citenamefont {Watanabe}, \citenamefont
  {Imai}, \citenamefont {Ishiyama}, \citenamefont {Jeong}, \citenamefont
  {Miyatake}, \citenamefont {Oyaizu}, \citenamefont {Matsuo}, \citenamefont
  {Sonoda},\ and\ \citenamefont {Wada}}]{HIR14}%
  \BibitemOpen
  \bibfield  {author} {\bibinfo {author} {\bibfnamefont {Y.}~\bibnamefont
  {Hirayama}}, \bibinfo {author} {\bibfnamefont {M.}~\bibnamefont {Mukai}},
  \bibinfo {author} {\bibfnamefont {Y.~X.}\ \bibnamefont {Watanabe}}, \bibinfo
  {author} {\bibfnamefont {N.}~\bibnamefont {Imai}}, \bibinfo {author}
  {\bibfnamefont {H.}~\bibnamefont {Ishiyama}}, \bibinfo {author}
  {\bibfnamefont {S.~C.}\ \bibnamefont {Jeong}}, \bibinfo {author}
  {\bibfnamefont {H.}~\bibnamefont {Miyatake}}, \bibinfo {author}
  {\bibfnamefont {M.}~\bibnamefont {Oyaizu}}, \bibinfo {author} {\bibfnamefont
  {Y.}~\bibnamefont {Matsuo}}, \bibinfo {author} {\bibfnamefont
  {T.}~\bibnamefont {Sonoda}}, \ and\ \bibinfo {author} {\bibfnamefont
  {M.}~\bibnamefont {Wada}},\ }\href@noop {} {\bibfield  {journal} {\bibinfo
  {journal} {J.\ Phys.\ B}\ }\textbf {\bibinfo {volume} {47}},\ \bibinfo
  {pages} {075201} (\bibinfo {year} {2014})}\BibitemShut {NoStop}%
\bibitem [{\citenamefont {Mukai}\ \emph {et~al.}(2015)\citenamefont {Mukai},
  \citenamefont {Hirayama}, \citenamefont {Imai}, \citenamefont {Ishiyama},
  \citenamefont {Jeong}, \citenamefont {Miyatake}, \citenamefont {Oyaizu},
  \citenamefont {Watanabe}, \citenamefont {Kim},\ and\ \citenamefont
  {Kimura}}]{MUK15}%
  \BibitemOpen
  \bibfield  {author} {\bibinfo {author} {\bibfnamefont {M.}~\bibnamefont
  {Mukai}}, \bibinfo {author} {\bibfnamefont {Y.}~\bibnamefont {Hirayama}},
  \bibinfo {author} {\bibfnamefont {N.}~\bibnamefont {Imai}}, \bibinfo {author}
  {\bibfnamefont {H.}~\bibnamefont {Ishiyama}}, \bibinfo {author}
  {\bibfnamefont {S.~C.}\ \bibnamefont {Jeong}}, \bibinfo {author}
  {\bibfnamefont {H.}~\bibnamefont {Miyatake}}, \bibinfo {author}
  {\bibfnamefont {M.}~\bibnamefont {Oyaizu}}, \bibinfo {author} {\bibfnamefont
  {Y.~X.}\ \bibnamefont {Watanabe}}, \bibinfo {author} {\bibfnamefont {Y.~H.}\
  \bibnamefont {Kim}}, \ and\ \bibinfo {author} {\bibfnamefont
  {S.}~\bibnamefont {Kimura}},\ }\href@noop {} {\bibfield  {journal} {\bibinfo
  {journal} {JPS.\ Conf.\ Proc.}\ }\textbf {\bibinfo {volume} {6}},\ \bibinfo
  {pages} {030127} (\bibinfo {year} {2015})}\BibitemShut {NoStop}%
\bibitem [{\citenamefont {Mukai}\ \emph {et~al.}(2016)\citenamefont {Mukai},
  \citenamefont {Hirayama}, \citenamefont {Ishiyama}, \citenamefont {Jung},
  \citenamefont {Miyatake}, \citenamefont {Oyaizu}, \citenamefont {Watanabe},
  \citenamefont {Kimura}, \citenamefont {Ozawa}, \citenamefont {Jeong},\ and\
  \citenamefont {Sonoda}}]{MUK16}%
  \BibitemOpen
  \bibfield  {author} {\bibinfo {author} {\bibfnamefont {M.}~\bibnamefont
  {Mukai}}, \bibinfo {author} {\bibfnamefont {Y.}~\bibnamefont {Hirayama}},
  \bibinfo {author} {\bibfnamefont {H.}~\bibnamefont {Ishiyama}}, \bibinfo
  {author} {\bibfnamefont {H.~S.}\ \bibnamefont {Jung}}, \bibinfo {author}
  {\bibfnamefont {H.}~\bibnamefont {Miyatake}}, \bibinfo {author}
  {\bibfnamefont {M.}~\bibnamefont {Oyaizu}}, \bibinfo {author} {\bibfnamefont
  {Y.~X.}\ \bibnamefont {Watanabe}}, \bibinfo {author} {\bibfnamefont
  {S.}~\bibnamefont {Kimura}}, \bibinfo {author} {\bibfnamefont
  {A.}~\bibnamefont {Ozawa}}, \bibinfo {author} {\bibfnamefont {S.~C.}\
  \bibnamefont {Jeong}}, \ and\ \bibinfo {author} {\bibfnamefont
  {T.}~\bibnamefont {Sonoda}},\ }\href@noop {} {\bibfield  {journal} {\bibinfo
  {journal} {Nucl.\ Instrum.\ Methods\ Phys.\ Res.\ B}\ }\textbf {\bibinfo
  {volume} {376}},\ \bibinfo {pages} {73} (\bibinfo {year} {2016})}\BibitemShut
  {NoStop}%
\bibitem [{\citenamefont {Firestone}(1996)}]{TOI}%
  \BibitemOpen
  \bibfield  {author} {\bibinfo {author} {\bibfnamefont {R.~B.}\ \bibnamefont
  {Firestone}},\ }\href@noop {} {\emph {\bibinfo {title} {Table of Isotopes}}}\
  (\bibinfo  {publisher} {John Wiley \& Sons, Inc.},\ \bibinfo {year}
  {1996})\BibitemShut {NoStop}%
\bibitem [{\citenamefont {Lee}\ \emph {et~al.}(1988)\citenamefont {Lee},
  \citenamefont {Savard}, \citenamefont {Crawford}, \citenamefont {Thekkadath},
  \citenamefont {Duong}, \citenamefont {Pinard}, \citenamefont {Liberman},
  \citenamefont {Blanc}, \citenamefont {Kilcher}, \citenamefont {Obert},
  \citenamefont {Oms}, \citenamefont {Putaux}, \citenamefont {Roussi$\grave{\rm
  e}$re},\ and\ \citenamefont {Sauvage}}]{LEE88}%
  \BibitemOpen
  \bibfield  {author} {\bibinfo {author} {\bibfnamefont {J.~K.~P.}\
  \bibnamefont {Lee}}, \bibinfo {author} {\bibfnamefont {G.}~\bibnamefont
  {Savard}}, \bibinfo {author} {\bibfnamefont {J.~E.}\ \bibnamefont
  {Crawford}}, \bibinfo {author} {\bibfnamefont {G.}~\bibnamefont
  {Thekkadath}}, \bibinfo {author} {\bibfnamefont {H.~T.}\ \bibnamefont
  {Duong}}, \bibinfo {author} {\bibfnamefont {J.}~\bibnamefont {Pinard}},
  \bibinfo {author} {\bibfnamefont {S.}~\bibnamefont {Liberman}}, \bibinfo
  {author} {\bibfnamefont {F.~L.}\ \bibnamefont {Blanc}}, \bibinfo {author}
  {\bibfnamefont {P.}~\bibnamefont {Kilcher}}, \bibinfo {author} {\bibfnamefont
  {J.}~\bibnamefont {Obert}}, \bibinfo {author} {\bibfnamefont
  {J.}~\bibnamefont {Oms}}, \bibinfo {author} {\bibfnamefont {J.~C.}\
  \bibnamefont {Putaux}}, \bibinfo {author} {\bibfnamefont {B.}~\bibnamefont
  {Roussi$\grave{\rm e}$re}}, \ and\ \bibinfo {author} {\bibfnamefont
  {J.}~\bibnamefont {Sauvage}},\ }\href@noop {} {\bibfield  {journal} {\bibinfo
   {journal} {Phys.\ Rev.\ C}\ }\textbf {\bibinfo {volume} {38}},\ \bibinfo
  {pages} {2985} (\bibinfo {year} {1988})}\BibitemShut {NoStop}%
\bibitem [{\citenamefont {Duong}\ \emph {et~al.}(1989)\citenamefont {Duong},
  \citenamefont {Pinard}, \citenamefont {Liberman}, \citenamefont {Savard},
  \citenamefont {Lee}, \citenamefont {Crawford}, \citenamefont {Thekkadath},
  \citenamefont {Blanc}, \citenamefont {Kilcher}, \citenamefont {Obert},
  \citenamefont {Oms}, \citenamefont {Putaux}, \citenamefont {Roussi$\grave{\rm
  e}$re}, \citenamefont {Sauvage},\ and\ \citenamefont {the
  ISOCELE~Collaboration}}]{DUO89}%
  \BibitemOpen
  \bibfield  {author} {\bibinfo {author} {\bibfnamefont {H.~T.}\ \bibnamefont
  {Duong}}, \bibinfo {author} {\bibfnamefont {J.}~\bibnamefont {Pinard}},
  \bibinfo {author} {\bibfnamefont {S.}~\bibnamefont {Liberman}}, \bibinfo
  {author} {\bibfnamefont {G.}~\bibnamefont {Savard}}, \bibinfo {author}
  {\bibfnamefont {J.~K.~P.}\ \bibnamefont {Lee}}, \bibinfo {author}
  {\bibfnamefont {J.~E.}\ \bibnamefont {Crawford}}, \bibinfo {author}
  {\bibfnamefont {G.}~\bibnamefont {Thekkadath}}, \bibinfo {author}
  {\bibfnamefont {F.~L.}\ \bibnamefont {Blanc}}, \bibinfo {author}
  {\bibfnamefont {P.}~\bibnamefont {Kilcher}}, \bibinfo {author} {\bibfnamefont
  {J.}~\bibnamefont {Obert}}, \bibinfo {author} {\bibfnamefont
  {J.}~\bibnamefont {Oms}}, \bibinfo {author} {\bibfnamefont {J.~C.}\
  \bibnamefont {Putaux}}, \bibinfo {author} {\bibfnamefont {B.}~\bibnamefont
  {Roussi$\grave{\rm e}$re}}, \bibinfo {author} {\bibfnamefont
  {J.}~\bibnamefont {Sauvage}}, \ and\ \bibinfo {author} {\bibnamefont {the
  ISOCELE~Collaboration}},\ }\href@noop {} {\bibfield  {journal} {\bibinfo
  {journal} {Phys.\ Lett.\ B}\ }\textbf {\bibinfo {volume} {217}},\ \bibinfo
  {pages} {401} (\bibinfo {year} {1989})}\BibitemShut {NoStop}%
\bibitem [{\citenamefont {Hilberath}\ \emph {et~al.}(1992)\citenamefont
  {Hilberath}, \citenamefont {Becker}, \citenamefont {Bollen}, \citenamefont
  {Kluge}, \citenamefont {Kr$\ddot{\rm o}$nert}, \citenamefont {Passler},
  \citenamefont {Rikovska}, \citenamefont {Wyss},\ and\ \citenamefont {the
  ISOLDE~Collaboration}}]{HIL92}%
  \BibitemOpen
  \bibfield  {author} {\bibinfo {author} {\bibfnamefont {T.}~\bibnamefont
  {Hilberath}}, \bibinfo {author} {\bibfnamefont {S.}~\bibnamefont {Becker}},
  \bibinfo {author} {\bibfnamefont {G.}~\bibnamefont {Bollen}}, \bibinfo
  {author} {\bibfnamefont {H.-J.}\ \bibnamefont {Kluge}}, \bibinfo {author}
  {\bibfnamefont {U.}~\bibnamefont {Kr$\ddot{\rm o}$nert}}, \bibinfo {author}
  {\bibfnamefont {G.}~\bibnamefont {Passler}}, \bibinfo {author} {\bibfnamefont
  {J.}~\bibnamefont {Rikovska}}, \bibinfo {author} {\bibfnamefont
  {R.}~\bibnamefont {Wyss}}, \ and\ \bibinfo {author} {\bibnamefont {the
  ISOLDE~Collaboration}},\ }\href@noop {} {\bibfield  {journal} {\bibinfo
  {journal} {Z.\ Phys.\ A -- Hadrons and Nuclei}\ }\textbf {\bibinfo {volume}
  {342}},\ \bibinfo {pages} {1} (\bibinfo {year} {1992})}\BibitemShut {NoStop}%
\bibitem [{\citenamefont {Kilcher}\ \emph {et~al.}(1992)\citenamefont
  {Kilcher}, \citenamefont {Putaux}, \citenamefont {Crawford}, \citenamefont
  {Dautet}, \citenamefont {Duong}, \citenamefont {Blanc}, \citenamefont {Lee},
  \citenamefont {Obert}, \citenamefont {Oms}, \citenamefont {Pinard},
  \citenamefont {Roussi$\grave{\rm e}$re}, \citenamefont {Sauvage},
  \citenamefont {Savard},\ and\ \citenamefont {Thekkadath}}]{KIL92}%
  \BibitemOpen
  \bibfield  {author} {\bibinfo {author} {\bibfnamefont {P.}~\bibnamefont
  {Kilcher}}, \bibinfo {author} {\bibfnamefont {J.~C.}\ \bibnamefont {Putaux}},
  \bibinfo {author} {\bibfnamefont {J.~E.}\ \bibnamefont {Crawford}}, \bibinfo
  {author} {\bibfnamefont {H.}~\bibnamefont {Dautet}}, \bibinfo {author}
  {\bibfnamefont {H.~T.}\ \bibnamefont {Duong}}, \bibinfo {author}
  {\bibfnamefont {F.~L.}\ \bibnamefont {Blanc}}, \bibinfo {author}
  {\bibfnamefont {J.~K.~P.}\ \bibnamefont {Lee}}, \bibinfo {author}
  {\bibfnamefont {J.}~\bibnamefont {Obert}}, \bibinfo {author} {\bibfnamefont
  {J.}~\bibnamefont {Oms}}, \bibinfo {author} {\bibfnamefont {J.}~\bibnamefont
  {Pinard}}, \bibinfo {author} {\bibfnamefont {B.}~\bibnamefont
  {Roussi$\grave{\rm e}$re}}, \bibinfo {author} {\bibfnamefont
  {J.}~\bibnamefont {Sauvage}}, \bibinfo {author} {\bibfnamefont
  {G.}~\bibnamefont {Savard}}, \ and\ \bibinfo {author} {\bibfnamefont
  {G.}~\bibnamefont {Thekkadath}},\ }\href@noop {} {\bibfield  {journal}
  {\bibinfo  {journal} {Nucl.\ Instrum.\ Methods\ Phys.\ Res.\ B}\ }\textbf
  {\bibinfo {volume} {70}},\ \bibinfo {pages} {537} (\bibinfo {year}
  {1992})}\BibitemShut {NoStop}%
\bibitem [{\citenamefont {Blanc}\ \emph {et~al.}(1999)\citenamefont {Blanc},
  \citenamefont {Lunney}, \citenamefont {Obert}, \citenamefont {Oms},
  \citenamefont {Putaux}, \citenamefont {Roussi$\grave{\rm e}$re},
  \citenamefont {Sauvage}, \citenamefont {Zemlyanoi}, \citenamefont {Pinard},
  \citenamefont {Cabaret}, \citenamefont {Duong}, \citenamefont {Huber},
  \citenamefont {Krieg}, \citenamefont {Sebastian}, \citenamefont {adn
  S.~P$\grave{\rm e}$ru},\ and\ \citenamefont {Genevey}}]{BLA99}%
  \BibitemOpen
  \bibfield  {author} {\bibinfo {author} {\bibfnamefont {F.~L.}\ \bibnamefont
  {Blanc}}, \bibinfo {author} {\bibfnamefont {D.}~\bibnamefont {Lunney}},
  \bibinfo {author} {\bibfnamefont {J.}~\bibnamefont {Obert}}, \bibinfo
  {author} {\bibfnamefont {J.}~\bibnamefont {Oms}}, \bibinfo {author}
  {\bibfnamefont {J.~C.}\ \bibnamefont {Putaux}}, \bibinfo {author}
  {\bibfnamefont {B.}~\bibnamefont {Roussi$\grave{\rm e}$re}}, \bibinfo
  {author} {\bibfnamefont {J.}~\bibnamefont {Sauvage}}, \bibinfo {author}
  {\bibfnamefont {S.}~\bibnamefont {Zemlyanoi}}, \bibinfo {author}
  {\bibfnamefont {J.}~\bibnamefont {Pinard}}, \bibinfo {author} {\bibfnamefont
  {L.}~\bibnamefont {Cabaret}}, \bibinfo {author} {\bibfnamefont {H.~T.}\
  \bibnamefont {Duong}}, \bibinfo {author} {\bibfnamefont {G.}~\bibnamefont
  {Huber}}, \bibinfo {author} {\bibfnamefont {M.}~\bibnamefont {Krieg}},
  \bibinfo {author} {\bibfnamefont {V.}~\bibnamefont {Sebastian}}, \bibinfo
  {author} {\bibfnamefont {M.~G.}\ \bibnamefont {adn S.~P$\grave{\rm e}$ru}}, \
  and\ \bibinfo {author} {\bibfnamefont {J.}~\bibnamefont {Genevey}},\
  }\href@noop {} {\bibfield  {journal} {\bibinfo  {journal} {Phys.\ Rev.\ C}\
  }\textbf {\bibinfo {volume} {60}},\ \bibinfo {pages} {054310} (\bibinfo
  {year} {1999})}\BibitemShut {NoStop}%
\bibitem [{\citenamefont {Witte1}\ \emph {et~al.}(2007)\citenamefont {Witte1},
  \citenamefont {Andreyev}, \citenamefont {Barr$\grave{\rm e}$}, \citenamefont
  {Bender}, \citenamefont {Cocolios}, \citenamefont {Dean}, \citenamefont
  {Fedorov}, \citenamefont {Fedoseyev}, \citenamefont {Fraile}, \citenamefont
  {Franchoo}, \citenamefont {Hellemans}, \citenamefont {Heenen}, \citenamefont
  {Heyde}, \citenamefont {Huber}, \citenamefont {Huyse}, \citenamefont
  {Jeppessen}, \citenamefont {K$\ddot{\rm o}$ster}, \citenamefont {Kunz},
  \citenamefont {Lesher}, \citenamefont {Marsh}, \citenamefont {Mukha},
  \citenamefont {Roussi$\grave{\rm e}$re}, \citenamefont {Sauvage},
  \citenamefont {Seliverstov}, \citenamefont {Stefanescu}, \citenamefont
  {Tengborn}, \citenamefont {de~Vel1}, \citenamefont {de~Walle}, \citenamefont
  {Duppen},\ and\ \citenamefont {Volkov}}]{WIT07}%
  \BibitemOpen
  \bibfield  {author} {\bibinfo {author} {\bibfnamefont {H.~D.}\ \bibnamefont
  {Witte1}}, \bibinfo {author} {\bibfnamefont {A.~N.}\ \bibnamefont
  {Andreyev}}, \bibinfo {author} {\bibfnamefont {N.}~\bibnamefont
  {Barr$\grave{\rm e}$}}, \bibinfo {author} {\bibfnamefont {M.}~\bibnamefont
  {Bender}}, \bibinfo {author} {\bibfnamefont {T.~E.}\ \bibnamefont
  {Cocolios}}, \bibinfo {author} {\bibfnamefont {S.}~\bibnamefont {Dean}},
  \bibinfo {author} {\bibfnamefont {D.}~\bibnamefont {Fedorov}}, \bibinfo
  {author} {\bibfnamefont {V.~N.}\ \bibnamefont {Fedoseyev}}, \bibinfo {author}
  {\bibfnamefont {L.~M.}\ \bibnamefont {Fraile}}, \bibinfo {author}
  {\bibfnamefont {S.}~\bibnamefont {Franchoo}}, \bibinfo {author}
  {\bibfnamefont {V.}~\bibnamefont {Hellemans}}, \bibinfo {author}
  {\bibfnamefont {P.~H.}\ \bibnamefont {Heenen}}, \bibinfo {author}
  {\bibfnamefont {K.}~\bibnamefont {Heyde}}, \bibinfo {author} {\bibfnamefont
  {G.}~\bibnamefont {Huber}}, \bibinfo {author} {\bibfnamefont
  {M.}~\bibnamefont {Huyse}}, \bibinfo {author} {\bibfnamefont
  {H.}~\bibnamefont {Jeppessen}}, \bibinfo {author} {\bibfnamefont
  {U.}~\bibnamefont {K$\ddot{\rm o}$ster}}, \bibinfo {author} {\bibfnamefont
  {P.}~\bibnamefont {Kunz}}, \bibinfo {author} {\bibfnamefont {S.~R.}\
  \bibnamefont {Lesher}}, \bibinfo {author} {\bibfnamefont {B.~A.}\
  \bibnamefont {Marsh}}, \bibinfo {author} {\bibfnamefont {I.}~\bibnamefont
  {Mukha}}, \bibinfo {author} {\bibfnamefont {B.}~\bibnamefont
  {Roussi$\grave{\rm e}$re}}, \bibinfo {author} {\bibfnamefont
  {J.}~\bibnamefont {Sauvage}}, \bibinfo {author} {\bibfnamefont
  {M.}~\bibnamefont {Seliverstov}}, \bibinfo {author} {\bibfnamefont
  {I.}~\bibnamefont {Stefanescu}}, \bibinfo {author} {\bibfnamefont
  {E.}~\bibnamefont {Tengborn}}, \bibinfo {author} {\bibfnamefont {K.~V.}\
  \bibnamefont {de~Vel1}}, \bibinfo {author} {\bibfnamefont {J.~V.}\
  \bibnamefont {de~Walle}}, \bibinfo {author} {\bibfnamefont {P.~V.}\
  \bibnamefont {Duppen}}, \ and\ \bibinfo {author} {\bibfnamefont
  {Y.}~\bibnamefont {Volkov}},\ }\href@noop {} {\bibfield  {journal} {\bibinfo
  {journal} {Phys.\ Rev.\ Lett.}\ }\textbf {\bibinfo {volume} {98}},\ \bibinfo
  {pages} {112502} (\bibinfo {year} {2007})}\BibitemShut {NoStop}%
\bibitem [{\citenamefont {Myers}\ and\ \citenamefont {Schmidt}(1983)}]{WIL83}%
  \BibitemOpen
  \bibfield  {author} {\bibinfo {author} {\bibfnamefont {W.~D.}\ \bibnamefont
  {Myers}}\ and\ \bibinfo {author} {\bibfnamefont {K.-H.}\ \bibnamefont
  {Schmidt}},\ }\href@noop {} {\bibfield  {journal} {\bibinfo  {journal}
  {Nucl.\ Phys.\ A}\ }\textbf {\bibinfo {volume} {410}},\ \bibinfo {pages} {61}
  (\bibinfo {year} {1983})}\BibitemShut {NoStop}%
\bibitem [{\citenamefont {Kimura}\ \emph {et~al.}(2016)\citenamefont {Kimura},
  \citenamefont {Ishiyama}, \citenamefont {Miyatake}, \citenamefont {Hirayama},
  \citenamefont {Watanabe}, \citenamefont {Jung}, \citenamefont {Oyaizu},
  \citenamefont {Mukai}, \citenamefont {Jeong},\ and\ \citenamefont
  {Ozawa}}]{KIM16}%
  \BibitemOpen
  \bibfield  {author} {\bibinfo {author} {\bibfnamefont {S.}~\bibnamefont
  {Kimura}}, \bibinfo {author} {\bibfnamefont {H.}~\bibnamefont {Ishiyama}},
  \bibinfo {author} {\bibfnamefont {H.}~\bibnamefont {Miyatake}}, \bibinfo
  {author} {\bibfnamefont {Y.}~\bibnamefont {Hirayama}}, \bibinfo {author}
  {\bibfnamefont {Y.~X.}\ \bibnamefont {Watanabe}}, \bibinfo {author}
  {\bibfnamefont {H.~S.}\ \bibnamefont {Jung}}, \bibinfo {author}
  {\bibfnamefont {M.}~\bibnamefont {Oyaizu}}, \bibinfo {author} {\bibfnamefont
  {M.}~\bibnamefont {Mukai}}, \bibinfo {author} {\bibfnamefont {S.~C.}\
  \bibnamefont {Jeong}}, \ and\ \bibinfo {author} {\bibfnamefont
  {A.}~\bibnamefont {Ozawa}},\ }\href@noop {} {\bibfield  {journal} {\bibinfo
  {journal} {Nucl.\ Instrum.\ Methods\ Phys.\ Res.\ B}\ }\textbf {\bibinfo
  {volume} {376}},\ \bibinfo {pages} {338} (\bibinfo {year}
  {2016})}\BibitemShut {NoStop}%
\bibitem [{MIN()}]{MIN94}%
  \BibitemOpen
  \href@noop {} {}\bibinfo {type} {{CERN Program Library entry
  D506}}\BibitemShut {NoStop}%
\bibitem [{\citenamefont {B$\ddot{\rm u}$ttgenbach}\ \emph
  {et~al.}(1984)\citenamefont {B$\ddot{\rm u}$ttgenbach}, \citenamefont
  {Glaeser}, \citenamefont {Roski},\ and\ \citenamefont {Tr$\ddot{\rm
  a}$ber}}]{BUT84}%
  \BibitemOpen
  \bibfield  {author} {\bibinfo {author} {\bibfnamefont {S.}~\bibnamefont
  {B$\ddot{\rm u}$ttgenbach}}, \bibinfo {author} {\bibfnamefont
  {N.}~\bibnamefont {Glaeser}}, \bibinfo {author} {\bibfnamefont
  {B.}~\bibnamefont {Roski}}, \ and\ \bibinfo {author} {\bibfnamefont
  {R.}~\bibnamefont {Tr$\ddot{\rm a}$ber}},\ }\href@noop {} {\bibfield
  {journal} {\bibinfo  {journal} {Z.\ Phys.\ A -- Atoms and Nuclei}\ }\textbf
  {\bibinfo {volume} {317}},\ \bibinfo {pages} {237} (\bibinfo {year}
  {1984})}\BibitemShut {NoStop}%
\bibitem [{\citenamefont {K$\ddot{\rm o}$ster}\ \emph
  {et~al.}(2011)\citenamefont {K$\ddot{\rm o}$ster}, \citenamefont {Stone},
  \citenamefont {Flanagan}, \citenamefont {Stone}, \citenamefont {Fedosseev},
  \citenamefont {Kratz}, \citenamefont {Marsh}, \citenamefont {Materna},
  \citenamefont {Mathieu}, \citenamefont {Mokkanov}, \citenamefont
  {Seliverstov}, \citenamefont {Serot}, \citenamefont {Sj$\ddot{\rm o}$din},\
  and\ \citenamefont {Volkov}}]{KOS11}%
  \BibitemOpen
  \bibfield  {author} {\bibinfo {author} {\bibfnamefont {U.}~\bibnamefont
  {K$\ddot{\rm o}$ster}}, \bibinfo {author} {\bibfnamefont {N.~J.}\
  \bibnamefont {Stone}}, \bibinfo {author} {\bibfnamefont {K.~T.}\ \bibnamefont
  {Flanagan}}, \bibinfo {author} {\bibfnamefont {J.~R.}\ \bibnamefont {Stone}},
  \bibinfo {author} {\bibfnamefont {V.~N.}\ \bibnamefont {Fedosseev}}, \bibinfo
  {author} {\bibfnamefont {K.~L.}\ \bibnamefont {Kratz}}, \bibinfo {author}
  {\bibfnamefont {B.~A.}\ \bibnamefont {Marsh}}, \bibinfo {author}
  {\bibfnamefont {T.}~\bibnamefont {Materna}}, \bibinfo {author} {\bibfnamefont
  {L.}~\bibnamefont {Mathieu}}, \bibinfo {author} {\bibfnamefont {P.~L.}\
  \bibnamefont {Mokkanov}}, \bibinfo {author} {\bibfnamefont {M.~D.}\
  \bibnamefont {Seliverstov}}, \bibinfo {author} {\bibfnamefont
  {O.}~\bibnamefont {Serot}}, \bibinfo {author} {\bibfnamefont {A.~M.}\
  \bibnamefont {Sj$\ddot{\rm o}$din}}, \ and\ \bibinfo {author} {\bibfnamefont
  {Y.~M.}\ \bibnamefont {Volkov}},\ }\href@noop {} {\bibfield  {journal}
  {\bibinfo  {journal} {Phys.\ Rev.\ C}\ }\textbf {\bibinfo {volume} {84}},\
  \bibinfo {pages} {034320} (\bibinfo {year} {2011})}\BibitemShut {NoStop}%
\bibitem [{\citenamefont {Wolbeck}\ and\ \citenamefont
  {Zioutas}(1972)}]{WOL72}%
  \BibitemOpen
  \bibfield  {author} {\bibinfo {author} {\bibfnamefont {B.}~\bibnamefont
  {Wolbeck}}\ and\ \bibinfo {author} {\bibfnamefont {K.}~\bibnamefont
  {Zioutas}},\ }\href@noop {} {\bibfield  {journal} {\bibinfo  {journal}
  {Nucl.\ Phys.\ A}\ }\textbf {\bibinfo {volume} {181}},\ \bibinfo {pages}
  {289} (\bibinfo {year} {1972})}\BibitemShut {NoStop}%
\bibitem [{\citenamefont {Saxena}\ and\ \citenamefont {Soares}(1981)}]{SAX81}%
  \BibitemOpen
  \bibfield  {author} {\bibinfo {author} {\bibfnamefont {R.~N.}\ \bibnamefont
  {Saxena}}\ and\ \bibinfo {author} {\bibfnamefont {J.~C.}\ \bibnamefont
  {Soares}},\ }\href@noop {} {\bibfield  {journal} {\bibinfo  {journal}
  {Hyperfine Interact.}\ }\textbf {\bibinfo {volume} {9}},\ \bibinfo {pages}
  {93} (\bibinfo {year} {1981})}\BibitemShut {NoStop}%
\bibitem [{\citenamefont {Krien}\ \emph {et~al.}(1977)\citenamefont {Krien},
  \citenamefont {Kroth}, \citenamefont {Saitovitch},\ and\ \citenamefont
  {Thomas}}]{KRI77}%
  \BibitemOpen
  \bibfield  {author} {\bibinfo {author} {\bibfnamefont {K.}~\bibnamefont
  {Krien}}, \bibinfo {author} {\bibfnamefont {K.}~\bibnamefont {Kroth}},
  \bibinfo {author} {\bibfnamefont {H.}~\bibnamefont {Saitovitch}}, \ and\
  \bibinfo {author} {\bibfnamefont {W.}~\bibnamefont {Thomas}},\ }\href@noop {}
  {\bibfield  {journal} {\bibinfo  {journal} {Z.\ Phys.\ A}\ }\textbf {\bibinfo
  {volume} {283}},\ \bibinfo {pages} {337} (\bibinfo {year}
  {1977})}\BibitemShut {NoStop}%
\bibitem [{\citenamefont {Blaum}\ \emph {et~al.}(2013)\citenamefont {Blaum},
  \citenamefont {Dilling},\ and\ \citenamefont {N$\ddot{\rm o}$rtersh$\ddot{\rm
  a}$user}}]{BLA13}%
  \BibitemOpen
  \bibfield  {author} {\bibinfo {author} {\bibfnamefont {K.}~\bibnamefont
  {Blaum}}, \bibinfo {author} {\bibfnamefont {J.}~\bibnamefont {Dilling}}, \
  and\ \bibinfo {author} {\bibfnamefont {W.}~\bibnamefont {N$\ddot{\rm
  o}$rtersh$\ddot{\rm a}$user}},\ }\href@noop {} {\bibfield  {journal}
  {\bibinfo  {journal} {Phys.\ Scr.\ T}\ }\textbf {\bibinfo {volume} {152}},\
  \bibinfo {pages} {014017} (\bibinfo {year} {2013})}\BibitemShut {NoStop}%
\bibitem [{\citenamefont {Ahmad}\ \emph {et~al.}(1985)\citenamefont {Ahmad},
  \citenamefont {Klempt}, \citenamefont {Ekstr$\ddot{\rm o}$m}, \citenamefont
  {Neugart}, \citenamefont {Wendt},\ and\ \citenamefont {the
  ISOLDE~collaboration}}]{AHM85}%
  \BibitemOpen
  \bibfield  {author} {\bibinfo {author} {\bibfnamefont {S.}~\bibnamefont
  {Ahmad}}, \bibinfo {author} {\bibfnamefont {W.}~\bibnamefont {Klempt}},
  \bibinfo {author} {\bibfnamefont {C.}~\bibnamefont {Ekstr$\ddot{\rm o}$m}},
  \bibinfo {author} {\bibfnamefont {R.}~\bibnamefont {Neugart}}, \bibinfo
  {author} {\bibfnamefont {K.}~\bibnamefont {Wendt}}, \ and\ \bibinfo {author}
  {\bibnamefont {the ISOLDE~collaboration}},\ }\href@noop {} {\bibfield
  {journal} {\bibinfo  {journal} {Z.\ Phys.\ A -- Atoms and Nuclei}\ }\textbf
  {\bibinfo {volume} {321}},\ \bibinfo {pages} {35} (\bibinfo {year}
  {1985})}\BibitemShut {NoStop}%
\bibitem [{\citenamefont {Raman}\ \emph {et~al.}(1989)\citenamefont {Raman},
  \citenamefont {Nestor}, \citenamefont {Kahane},\ and\ \citenamefont
  {Bhatt}}]{RAM89}%
  \BibitemOpen
  \bibfield  {author} {\bibinfo {author} {\bibfnamefont {S.}~\bibnamefont
  {Raman}}, \bibinfo {author} {\bibfnamefont {J.}~\bibnamefont {Nestor}},
  \bibinfo {author} {\bibfnamefont {S.}~\bibnamefont {Kahane}}, \ and\ \bibinfo
  {author} {\bibfnamefont {K.}~\bibnamefont {Bhatt}},\ }\href@noop {}
  {\bibfield  {journal} {\bibinfo  {journal} {At.\ Data\ Nucl.\ Data Tables}\
  }\textbf {\bibinfo {volume} {42}},\ \bibinfo {pages} {1} (\bibinfo {year}
  {1989})}\BibitemShut {NoStop}%
\end{thebibliography}%

\end{document}